\documentclass[
    aps,
    pra,
    reprint,
    longbibliography,
    nofootinbib,
    superscriptaddress,
    floatfix
]{revtex4-2}

\usepackage{epsfig,amsmath,amssymb,ragged2e}
\usepackage[outdir=./]{epstopdf}
\usepackage{comment}
\usepackage{hyperref}
\usepackage[english]{babel}
\babeladjust{autoload.bcp47 = on}

\newcommand{\Exp}[1]{\exp{\left(#1\right)}}
\newcommand{\expb}[1]{\exp{\left(#1\right)}}
\newcommand{\expbin}[1]{\exp{\bigl(#1\bigr)}}
\newcommand{\expbb}[1]{\exp{\left[#1\right]}}
\newcommand{\tr}[1]{{\rm Tr}\left(#1\right)}

\usepackage{gensymb} 

\usepackage[dvipsnames, table]{xcolor}
\usepackage{textcomp}
\usepackage{graphicx}
\usepackage[caption=false]{subfig}
\usepackage{array}

\usepackage[export]{adjustbox}

\definecolor{darkcyan}{rgb}{0.0, 0.55, 0.55}

\newcommand{\mycomment}[1]{}

\makeatletter

\renewcommand\onecolumngrid{
	\do@columngrid{one}{\@ne}%
	\def\set@footnotewidth{\onecolumngrid}
	\def\footnoterule{\kern-6pt\hrule width 1.5in\kern6pt}%
}

\renewcommand\twocolumngrid{
	\def\footnoterule{
		\dimen@\skip\footins\divide\dimen@\thr@@
		\kern-\dimen@\hrule width.5in\kern\dimen@}
	\do@columngrid{mlt}{\tw@}
}%

\def\@appendixcntformat#1{\csname the#1\endcsname}
\def\@hangfrom@appendix#1#2#3{%
  #1#2\quad\MakeTextUppercase{#3}%
}

\makeatother    

\begin{document}
	
	\title{Quantum Keynesian Beauty Contest with Continuous Variables}


\author{Hiu Ming Lau}
\affiliation{School of Science, Royal Melbourne Institute of Technology (RMIT University), Melbourne Victoria 3001, Australia}
\affiliation{Commonwealth Scientific and Industrial Research Organisation (CSIRO), Clayton Victoria 3168, Australia}
\author{Kareem Raslan}
\affiliation{Commonwealth Scientific and Industrial Research Organisation (CSIRO), Clayton Victoria 3168, Australia}
\affiliation{School of Electrical and Electronic Engineering, The University of Adelaide, Adelaide South Australia 5005, Australia}
\author{Kieran Hymas}
\affiliation{Commonwealth Scientific and Industrial Research Organisation (CSIRO), Clayton Victoria 3168, Australia}
\author{Qiang Sun}
\affiliation{School of Science, Royal Melbourne Institute of Technology (RMIT University), Melbourne Victoria 3001, Australia}
\author{Azhar~Iqbal}
\affiliation{School of Electrical and Electronic Engineering, The University of Adelaide, Adelaide South Australia 5005, Australia}
\author{Derek Abbott}
\affiliation{School of Electrical and Electronic Engineering, The University of Adelaide, Adelaide South Australia 5005, Australia}
\author{James Q. Quach}
\affiliation{Commonwealth Scientific and Industrial Research Organisation (CSIRO), Clayton Victoria 3168, Australia}
\author{Andrew D. Greentree}
\affiliation{School of Science, Royal Melbourne Institute of Technology (RMIT University), Melbourne Victoria 3001, Australia}
	\begin{abstract}
The development of quantum computing has opened up opportunities to re-explore conventional game models by the integration of quantum processes. Through the established framework of quantum game theory, we examine the problem of speculative pricing in the Keynesian Beauty Contest model -- the $p$-Beauty Contest ($p$-BC). In the $p$-BC, players are rewarded with the optimal payoff for successfully bidding at some fraction $p$ of the average of all of the players' bids. Consequently, \textit{rational} plays in the $p$-BC inexorably lead to boom or bust strategies for $p>1$ or $p<1$ respectively. We demonstrate a quantum continuous-variable implementation of $p$-BC where the boom-and-bust rational behavior is mitigated through a global transformation of the players' bids generated by increasing multimode squeezing. While we focus on a quantum mechanical implementation of the $p$-BC, we show that an entirely analogous game can be formulated via classical channels by a global mediator. Thus we identify global mediation as the underlying game-theoretic mechanism for quenching boom and bust phenomena in our model. We analyze the quantum $p$-BC for two, three, and many players in a mean-field type implementation. We derive an analytical form for the probability distribution of the payoff under single-value measurements of the quantum processing circuit, and explore how the rational behavior differs from the expected-value measurement setting.

	\end{abstract}
	\maketitle

\section{Introduction}
The Keynesian beauty contest (KBC) is a simplified analogy for stock market pricing, where investors choose their bids based on their perceptions of the likely strategies of the other investors \cite{Keynes_1937, Lanteri_2011, Allen_2006, Brockway_1983, Morris_2002}. The KBC involves a selection of judges tasked with ordering the best-looking contestants from a crowd, where the judge who selects the most popular contestant, as determined by the judges, wins. One quickly realizes selecting contestants based on one's own genuine opinion of beauty is a losing strategy. Instead, a better strategy is to rank the contestants according to the rankings that the other judges are likely to make. 

Inspired by the idea of KBC, numerous studies have investigated the nested rational process in speculative decision-making \cite{Satorra_2002, Nagel_2008}. Such experimental realizations of KBC have participants guessing some scaled average of all bids submitted by all players, where the  bidding bias introduced by the scaling factor is compounded by rational speculative strategies, which results in players bidding towards extreme values. Specifically, a popular setting of this game has been \textit{Guess $2/3$ of the Average}. To quantify the pattern of behaviors arising from the experiments, the $p$-Beauty Contest ($p$-BC) was then created as a quantitative game model to explicitly examine such interactions of participants, where $p$ is defined to be the scaling factor. This captured the rationalization mechanism explaining how speculative trading contributes to boom-and-bust behavior \cite{Ho_1998, Sbriglia_2008, Cartwright_2011}. 

The $p$-BC game highlights the nested logical assumptions of rational players in attempting to derive the optimal strategy. Under this symmetric game, rational players may assume other players are able to think of the same rational strategy as themselves. From here, a counter strategy can be derived for the given assumption, applying the assumption again with the newly derived counter strategy, then a new counter strategy is found, and this cycle of assumptions and counter strategy can continue on and on. Consequently, the continuing iterations lead rational players in a game of $p$-BC to bid towards the extremes of the allowed strategy space: for $p<1$, players bid towards the minimum, and for $p>1$ players bid towards the maximum. The consequences of such speculative process are reflective of the form of market bubbles and destabilization observed in the economy \cite{Hirota_2002, Morris_2002}.

Quantum game theory began by considering an expansion of conventional game models, where quantum processes are integrated into the player strategies \cite{Meyer_1999, Eisert_1999}. By exploring the rational behaviors between quantum actors, quantum games have demonstrated a quantum mechanical approach in exploring games by embedding the strategy space in a richer quantum state space \cite{Flitney_2009, Khan_2018, Sanz_2025}. While many quantum games utilize quantum operators and entanglement in discrete-variable quantum systems, continuous-variable quantum systems can be a natural and effective choice in expanding classical game models \cite{Li_2002, Zhou_2005, Guo_2008, Liao_2008, Yang_2024}. With the development of quantum computing, quantum games have been used to explore potential economic environments implemented on quantum hardware, sparking interest in designing quantum-based markets through the framework of quantum game theory in recent years \cite{Hymas_2025, Ikeda_2021, Ding_2025}. 

Collective or global structures are also familiar in classical game theory. In aggregative games, individual incentives depend on population-level quantities constructed from the complete strategy profile, while in beauty-contest settings common information can acquire an amplified strategic role because players respond both to the information itself and to the anticipated responses of others \cite{Morris_2002,AcemogluJensen_2013}. Global effects may also arise through explicit coordination between players; for example, Flitney and Greentree considered coalitions in the Minority Game, comparing classically coordinated strategies based on communication with strategies enabled by shared quantum entanglement \cite{FlitneyGreentree_2007}. These examples motivate a broader question explored here: whether the strategic consequences associated with a quantum collective process can be understood in terms of an equivalent global operation in a classical game.

Here, we present an $N$-player quantum version of the $p$-BC, where the player bids are implemented as single-value (SV) or expectation-value (EV) measurements of a quantum state. We first present the resultant rational behavior in the quantum $p$-BC under EV measurements to demonstrate the ability of the quantization circuit in mitigating boom-and-bust behaviors. We show that the multimode quantum circuit acts globally across the complete strategy profile, preserving the population mean while modifying each player's deviation from that mean. This transformation renormalizes the effective beauty-contest parameter, ultimately mitigating the iterative boom-and-bust behavior of the classical game. Through the classical process of the probability-weighted average, the EV quantum $p$-BC enables a reparameterization of the quantum payoff function which resembles the classical payoff. This treatment allows the application of established conventional game theoretic analysis, resulting in a clear description for the rational behavior in the quantum game. Importantly, the Gaussian quantum construction admits an exact quantum-inspired classical realization in which a global mediator applies the same mean-dependent transformation directly to all submitted bids. The quantum and classical mechanisms are physically distinct, but occupy the same architectural role: the strategic variables are transformed globally rather than independently player by player. Using the results of the EV game, the SV quantum $p$-BC can be explored with greater clarity under the added complexity of probabilistic payoff. We demonstrate that the mitigation of boom-and-bust is preserved in the SV game, while the probabilistic payoff gives rise to complex player interactions.

This paper is organized as follows: Section \ref{sec:Classical} introduces the structure and results of the classical $p$-BC game with the established game-theoretic analysis. Section \ref{sec:Quantum} presents the \textit{quantization scheme} and the EV quantum $p$-BC, where the EV quantum payoff is re-parameterized to apply conventional game-theoretic analysis and mean-field type analyses. Section \ref{sec:SV_game} will introduce the SV quantum $p$-BC, derive a general expression for the random payoff function, and explore the rational behavior under the probabilistic quantum payoff.

\subsection{Contributions}

The main results presented in this article are as follows:

\begin{enumerate}
    \raggedright
    \item\textbf{Continuous-variable quantum $p$-Beauty Contest to mitigate boom-and-bust behavior}
    
    We demonstrate a CV quantization process for the $p$-BC which mitigates the existing boom-and-bust behavior of the classical game. The strength of this effect is controlled by the squeezing parameter $\gamma$, and we show that the resulting global transformation of the player strategies admits an exact quantum-inspired classical interpretation.
    \item\textbf{Reparameterization of quantum payoff to apply conventional game theoretic analysis}
    
    \noindent Through a reparameterization of EV quantum $p$-BC payoff function, we apply the same game theoretic analysis of iterative elimination on the quantum game, providing a clear and comparable result to the classical game.
    \item\textbf{Mean-field type analysis for the $N$-player quantum $p$-Beauty Contest}
    
    We perform a mean-field type analysis on the EV quantum $p$-BC under different settings. Through a simplified $N$-player quantum game, we demonstrate the effect of increasing levels of entanglement $\gamma$ on the players' best responses. 
    \item\textbf{Game theoretic analysis with probabilistic quantum payoff}
    
    We derive an analytical expression for the pdf of the probabilistic payoff in SV quantum $p$-BC. We present the characteristics of the probabilistic SV payoff under varying discretization settings, including a divergence in its continuous probability density. We demonstrate the on-average optimal strategy of the probabilistic SV quantum payoff resembles the non-probabilistic EV quantum payoff. 
    
\end{enumerate}



\section{Classical $p$-Beauty Contest} \label{sec:Classical}

The $p$-BC is more commonly known as Guess $p$ of the Average, where $p \in (0, \infty)$ is some scalar multiple for the average. The $N$-player game is encapsulated in its payoff function \cite{Yildiz_2012}:
\begin{align}
    \$_i(\textbf{x}) = -\left(x_i - p \frac{\sum_{j=1}^N x_j}{N}\right)^2.
    \label{eqn:classical_payoff}
\end{align}

\noindent This is the payoff function of player $i$, with $x_i$ their input bid; $N$ the total number of players in the game; $\textbf{x} = (x_1, x_2, ...x_i..., x_N)$ the strategy profile of all players. The payoff is an inverted parabola to penalize bids far from $p$ of the average, with $0$ being the maximum payoff. 

\subsection{Iterative Elimination} \label{sec:quali_iterative}

The concept of $k$-\textit{rationality}, as derived from \textit{Iterative Deletion of Strictly Dominated Strategies} (IDSDS), is widely used in $p$-BC to describe how logical assumptions and speculative thinking leads rational players to bid towards the extremum of the strategy space \cite{Ho_1998, Nagel_2008, Cartwright_2011}. The $k$ represents the number of iterations the player goes through in predicting the behavior of other players. 

\begin{figure}[tb]
    \centering
    \includegraphics[width=\linewidth]{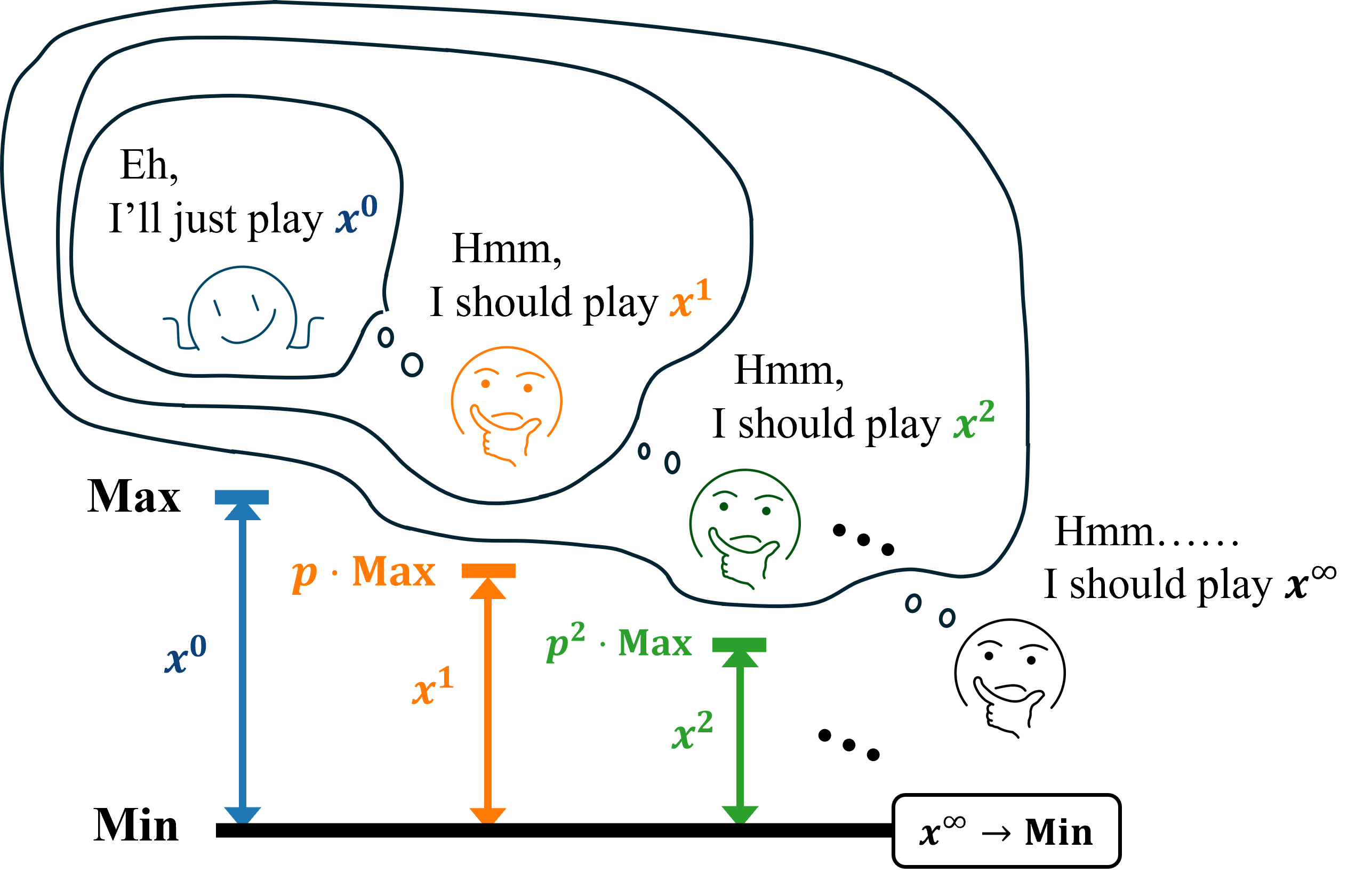}
    \caption{A visualization of the IDSDS process for ``Guess $2/3$ of the Average'', where each player chooses a number from $[0, 100]$. With increasing level of rationality, rational players are compelled to choose smaller and smaller numbers. The naive player chooses a number from the entire range $x^{k=0}=[0, 100]$; a $1$-rational player picks their number within a smaller range of $x^{k=1}=[0, 66.67]$; a $2$-rational player chooses within $x^{k=2}=[0, 44.44]$; and finally, the $\infty$-rational player only picks $0$, which is the \textit{Nash equilibrium} of this game.}
    \label{fig:23_bar}
\end{figure}

Here, we demonstrate the iterative process through an example game of ($p=2/3$)-BC, with valid bids between $0$ and $100$. Regardless of strategy, all players will bid within the strategy space $x_{j\neq i}^{k=0} = [0, 100]$, hence the maximum $2/3$ average is $2/3 \cdot 100\approx66.67$. The set of optimal strategies is then limited to $x_i^{k=1} = [0, 66.67]$. If a rational player assumes all other players know this, then their new optimal strategy is now $x_i^{k=2} = [0, 44.44]$ with the new maximum $2/3$ average $(2/3)^2 \cdot 100\approx44.44$. After $k$ steps, the optimal strategy is then $x_i^{k} = [0, (2/3)^k\cdot 100]$. For $k\rightarrow \infty$, the optimal strategy is simply the bust $x_i^{k\rightarrow \infty} = 0$, which is also the Nash equilibrium of the ($p=2/3$)-BC game \cite{Yildiz_2012}. This rational iterative process is visualized in Figure~(\ref{fig:23_bar}).

At every iteration, the rational player $i$ as the $k$-rational player assumes the crowd to be $(k-1)$-rational players, then the strictly decreasing region of player $i$'s payoff spectrum $\$_i(x_1, x_2,...x_i^{k}...,x_{N}) > \$_i(x_1, x_2,...x_{\rm dominated}^k...,x_{N})$ is found and eliminated to obtain the optimal strategy \cite{Yildiz_2012}. A quantitative IDSDS process is shown in Section~\ref{sec:Expected_IDSDS}, where the exact classical game-theoretic analysis is applied to the EV quantum $p$-BC.

\subsection{Nash Equilibria}

The movements of the players' bids are driven by the iterative elimination processes described above. Under the limiting assumption of infinitely rational players $k \rightarrow \infty$, this $k$-rational strategy results in players bidding at the respective Nash equilibrium, and game theoretically eliminates any subsequent deviating bids. 

For ($p<1$)-BC or \textit{Guessing Smaller than the Average}, the Nash equilibrium is \textit{bust} $\textbf{x}_{\rm Nash} = \{(0, 0, ..., 0)\}$. For \textit{Guessing Larger than the Average} ($p>1$)-BC, the relevant Nash equilibrium is \textit{boom} $\textbf{x}_{\rm Nash} = \{(\rm 100, 100, ..., 100)\}$. For \textit{Guessing the Average} ($p=1$)-BC, there are no dominated strategies and hence infinitely many Nash equilibria \cite{Nagel_1995} across the continuous strategy space. Significantly, given neither boom nor bust is the sole Nash equilibrium in $p=1$, rational strategies no longer result in extremum bids.

The $k$-rational strategy derived from single-round game analysis is a major tool in analyzing multi-round experimental $p$-BC by categorizing the players as different levels of $k$-rational \cite{Cartwright_2011, Nagel_2008, Sbriglia_2008}. The $p$-BC experiments showed how different $k$-rational players learn to bid towards the extremum Nash equilibria. Importantly, as $p$ deviates further from $1$, the players learn to bid towards the extremum faster.

In this paper, we will continue to use $p<1$ for mathematical clarity, while the same derivations hold equally true for $p>1$ with mirroring results. In following sections, we present the quantum game theoretic quantization and discuss how the quantum $p$-BC mitigates boom-and-bust behaviors.

\section{Continuous-variable Quantization} \label{sec:Quantum}

\begin{figure}[!tbh]
    \centering
    \includegraphics[width=\linewidth]{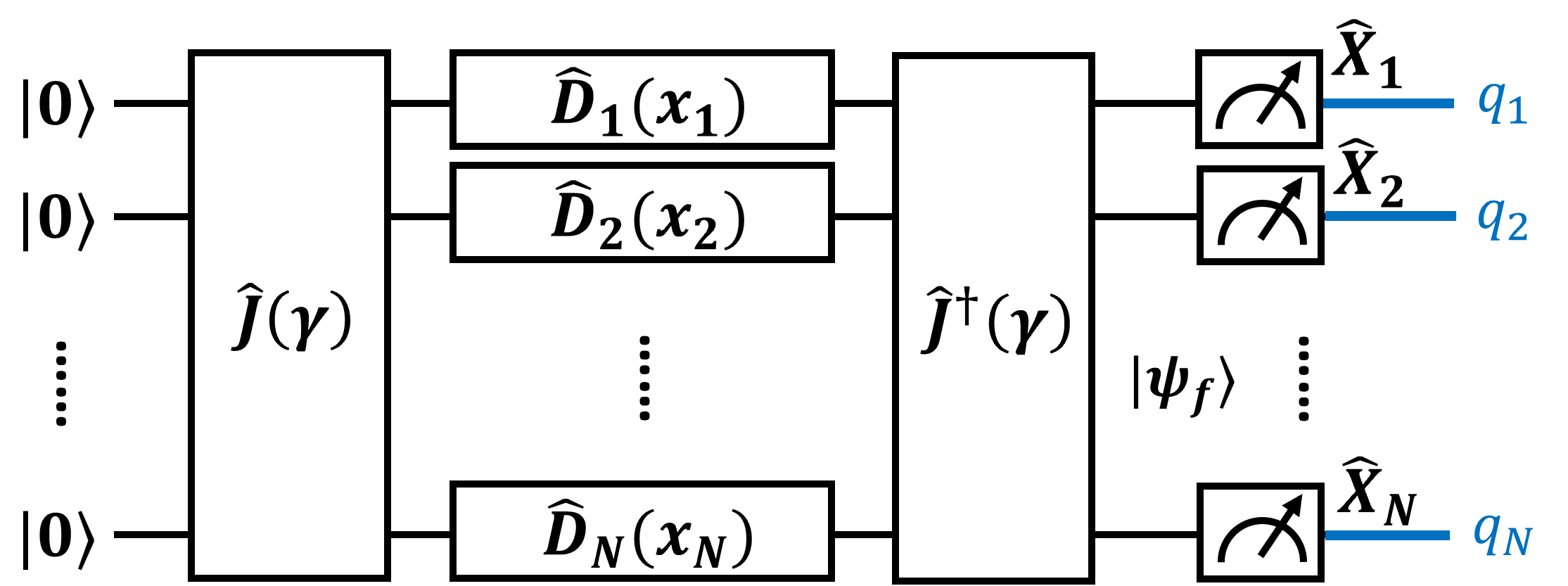}
    \caption{$N$-player continuous-variable quantization scheme. The core element in the quantization is this quantum circuit. Here, $\hat{D}_i$ is the single-mode displacement operator encoding player i's bid on their corresponding mode, where the displacement parameter $x_i$ is limited to the position axis, omitting the complex momentum component. The two $N$-mode squeezing operators $\hat{J}$ \& $\hat{J}^\dagger$ are as described in Equation~\eqref{eqn:J_op}, which introduce correlations between the modes and hence the strategies of all players. Finally, each mode is measured in its corresponding positional basis $\hat{X}_i$ given $\hat{D}_i(x_i)$, with the measurement results as $q_i$.}
    \label{fig:quant_circuit}
\end{figure}

Now we explore a quantum version of $p$-BC, where a quantum circuit introduces entanglement between continuous-variable player bids. Our $N$-player quantization scheme is an extension of the 2-player Li-Du-Massar quantization \cite{Li_2002} as shown in Figure~(\ref{fig:quant_circuit}). 

The present CV construction in the quantization process is Gaussian and hence classically simulable. We utilize this property to isolate a classical game-theoretic mechanism from the resultant quantized game, which employs multimode squeezing and intermediate entanglement. This process results in an equivalent quantum-inspired classical game, through which we characterize the strategic consequences of the quantum-derived global transformation. 

In this quantum $p$-BC, each player inputs their bid via their assigned single-mode positional displacement operators $\hat{D}_i = \expbin{-i x_i \hat{P}_i}$, where $\hat{P}_i = i/\sqrt{2}\ \bigl(\hat{a}_i^\dagger - \hat{a}_i\bigr)$ \cite{Lo_1993}. The imaginary component of the displacement $x_i$ is excluded, and the displacement operator $\hat{D}_i$ limited to the position basis $\hat{X}_i$. By performing quadrature measurements of the modes with respect to the positional basis $\hat{X}_i = 1/\sqrt{2}\bigl(\hat{a}_i^\dagger + \hat{a}\bigr)$, the quantization circuit produces a set of quantum-processed bids $\textbf{q}$. These measurement outcomes $\textbf{q}$ are substituted into Equation~\eqref{eqn:classical_payoff} to calculate every player's quantum payoff. From here, we can write the quantum payoff function as: 

\begin{align}
    \$_i^Q(\textbf{x}) = \$_i[\textbf{q}(\textbf{x})] = -\left(q_i - p \frac{\sum_{j=1}^N q_j}{N}\right)^2.
    \label{eqn:quant_payoff_q}
\end{align}

There are two $N$-mode squeezing operations $\hat{J}$ and $\hat{J}^\dagger$ which introduce entanglement between the modes and players' bids. The operations are described by \cite{Lo_1993, Lo_2003}: 
\begin{align}
    \hat{J}(\gamma) = \expb{-\gamma\sum\limits_{i,j=1}^N \left(\delta_{ij} - 1/N\right) \left(\hat{a}^\dagger_i\hat{a}^\dagger_j - \hat{a}_i\hat{a}_j\right)}
    \label{eqn:J_op}
\end{align}

\noindent where $\gamma$ represents the level of squeezing, and $\hat{J}^\dagger$ is obtained by reversing the sign of the exponent. 

Importantly, through this bid entangling process, each player's input $x_i$ can deviate from their own quantum-processed bid $q_i$ depending on all other players' bids. A rational player in the quantum game must not only consider the other players' inputs, but additionally account for intrinsic quantum measurement noise introduced by the quantization circuit. While in the classical game, a rational player can devise a strategy to bid a particular number, and their input will simply be their resultant bid. 

\subsection{Expected-Value Quantum
\texorpdfstring{$p$}{p}-Beauty Contest}
In the EV quantum $p$-BC, each set of player inputs is repeated over a large ensemble of experiments, with the expected quantum-processed bid $\langle \textbf{q}\rangle$ determining the outcome of the game. The expected quantum payoff for player $i$ is thus:
\begin{align}
    \langle\$_i^Q\rangle &= -\left(\langle q_i\rangle(\gamma) - p \frac{\sum_{j=1}^N \langle q_j\rangle(\gamma)}{N}\right)^2    
\end{align}
    
\noindent where we use $\langle\$_i^Q\rangle$ to denote the EV processed quantum payoff. This treatment significantly simplifies the game-theoretic analysis, and facilitates the derivation of the average rational behavior for the quantum $p$-BC.

To express $\langle \$_i^Q\rangle$ as a function of players' inputs $\textbf{x}$, we derive $\langle q_i\rangle(\textbf{x})$ by finding the final wave function in the quantization circuit, then the expectation value with respect to the positional basis $\hat{X_i}$:
\begin{gather}
\begin{aligned}
    \nonumber |\psi_f\rangle &= \hat{J}^\dagger\prod_k^N\hat{D}_k(x_k)\hat{J}|0\rangle^{\otimes N}\\
    &= \bigotimes_k^N \expb{-i\left[e^{2\gamma}(x_k-\bar{x})+\bar{x}\right]\hat{P}_k}|0\rangle_k
\end{aligned}\\
\begin{aligned}
    \langle q_i\rangle(\textbf{x}, \gamma) = \tr{\hat{\rho}_f\hat{X}_i} = e^{2\gamma}\left(x_i-\bar{x}\right)+\bar{x}\label{eqn:expected_qi}
\end{aligned}
\end{gather}

\noindent where $\bar{x}$ is the averaged player input $\bar{x} = \frac{1}{N}\sum_{i=1}^N x_i$; $\hat{\rho}_f = |\psi_f\rangle\langle\psi_f|$. 

The chosen squeezing operations and the displacement operators result in a separable state as shown in the above form of $|\psi_f\rangle$. Each mode is a coherent state with a positional displacement determined by the inputs from all players through the preceding multimode entangling transformation. The resulting global dependence of each processed bid on the complete strategy profile is controlled by the squeezing parameter $\gamma$.

Under the EV setting, the classical $p$-Beauty Contest model is fully retained as the quantum game under no entanglement $\langle q_i\rangle|_{\gamma = 0} = x_i$. Importantly, the average bid is also preserved, where the average EV quantum-processed bid equals the average player input $\langle\bar{q}\rangle = \bar{x}$. 

\subsection{Effect of Squeezing Operations} \label{subsec:Squeezing_Effect}

\begin{figure}[ht!]
    \centering
\subfloat[]{%
    \includegraphics[width=0.8\linewidth]{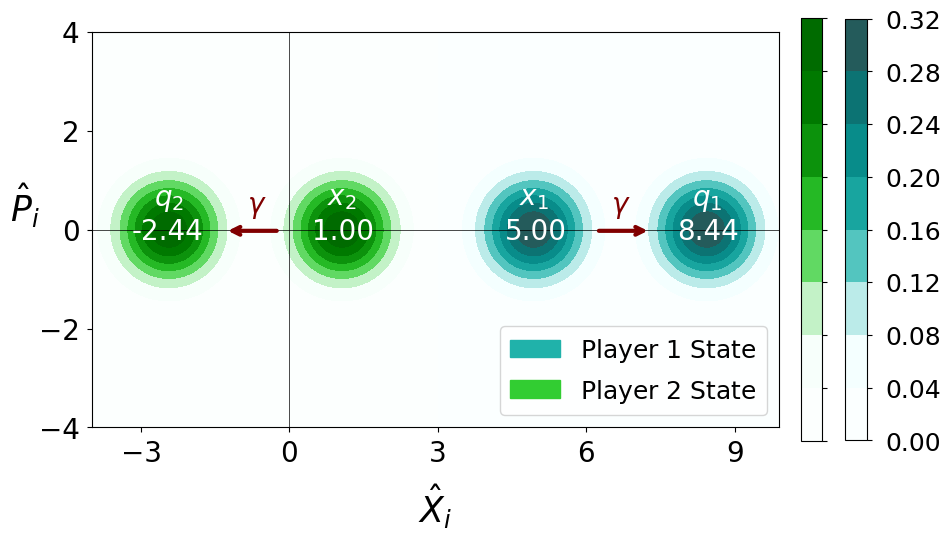}
}

    \centering
 \subfloat[]{%
    \includegraphics[width=0.85\linewidth]{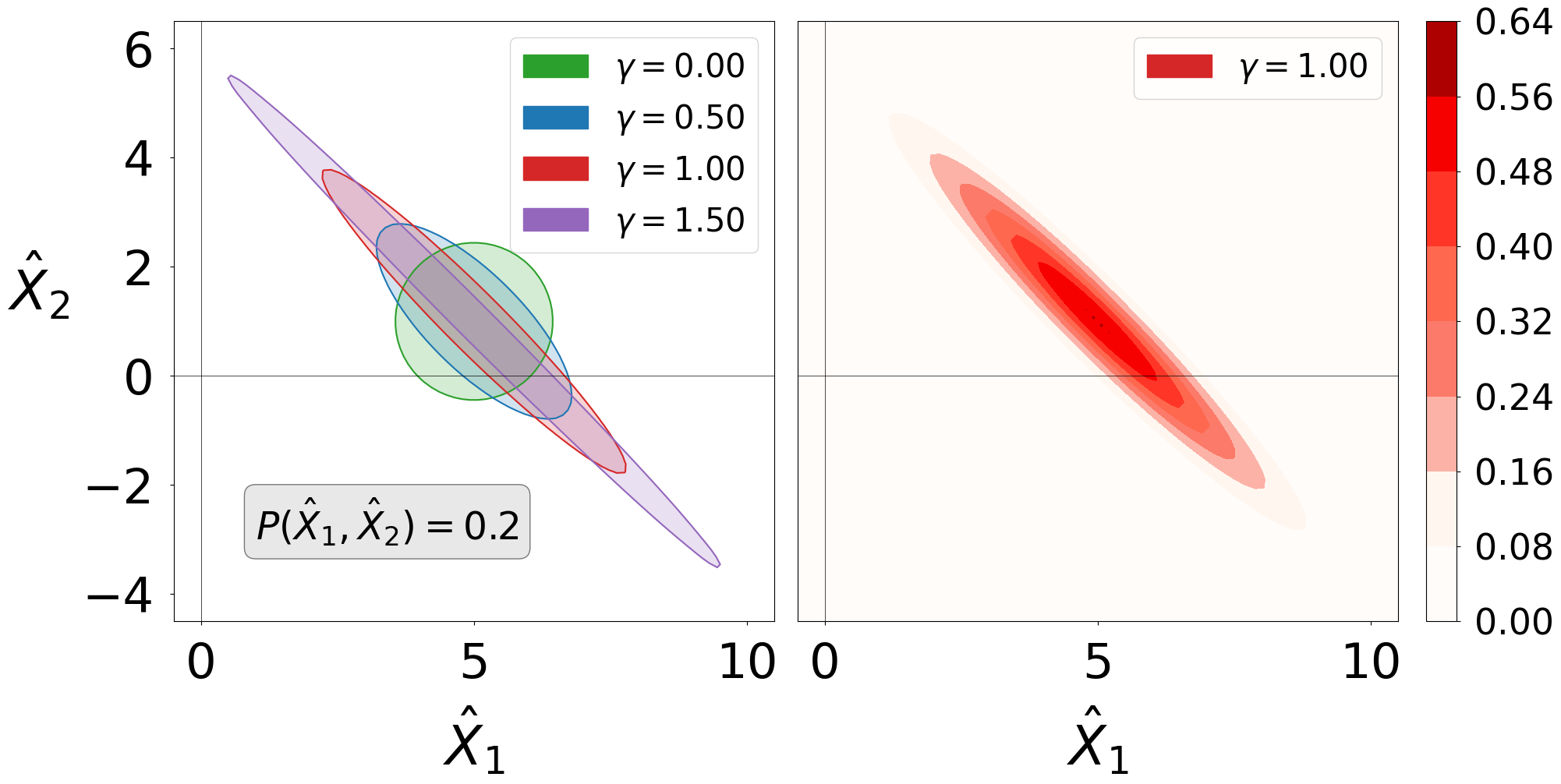}
}
    \caption{Wigner Function Plots. Given $2$ players who have input bids $x_1 = 5$ and $x_2 = 1$ through the quantization circuit respectively: a) For $\gamma=0$, there is no entanglement, and only the displacement operations $\hat{D}_i$ are in effect. The modes encoding each player's input are \textit{displaced coherent states} with positions corresponding to the inputs. With $\gamma=0.5$, the states of the modes remain displaced coherent states through the sandwiching gates $\hat{J}$ \& $\hat{J}^\dagger$ under non-zero squeezing, where the entanglement results in the correlation of the expected position $\langle \hat{X}_i\rangle$. b) On the other hand, if $\hat{J}$ and $\hat{D}_i$ are applied without $\hat{J}^\dagger$, the resultant modes are in an entangled state, where measurement outcomes of the modes are anti-correlated.}
    \label{fig:dist_funcs}
\end{figure}

The influence of $\gamma$ on the correlation of bids can be demonstrated through a $2$-player example of the EV quantum game. The quantum states across the circuit are presented in Figure~(\ref{fig:dist_funcs}). In this example, player $1$ and $2$ provide the inputs $x_1 = 5$, $x_2 = 1$, respectively, and the expected positional measurement $\hat{X}_1, \hat{X}_2$ for the quantization circuit output is simply the same as each player's input: $\langle q_1\rangle = x_1$, $\langle q_2\rangle = x_2$. Under zero squeezing ($\gamma=0$), the displacement operators $\hat{D}_1(x_1)$ \& $\hat{D}_2(x_2)$ simply encodes the classical game as shown in Figure~(\ref{fig:dist_funcs}a). 

With the addition of the first squeezing operator $\hat{J}$ for $\gamma > 0$ to the displacement operators, the quantization circuit creates a two-mode displaced squeezed state, where the positional measurements of each mode become anti-correlated as depicted in Figure~(\ref{fig:dist_funcs}b). For example, if a larger than average $q_1 > x_1$ measurement is obtained for player $1$'s mode, then there is a high probability of obtaining a smaller than average $q_2 < x_2$ for player $2$. The effect is amplified by increased level of squeezing $\gamma \uparrow$. 

By adding the second squeezing operation $\hat{J}^\dagger$, the anti-correlation in the positional measurement now translates to anti-correlation in the positional displacement of the coherent states as shown in Figure~(\ref{fig:dist_funcs}a). The effect of this entanglement configuration on the bids is best described by Equation~\eqref{eqn:expected_qi}, where the expected distance between each player's inputs and the average input is amplified by the squeezing parameter $\gamma$. For example, a player who bids away from the average may find their quantum-processed bids ending up further away. 

In the following, we utilize the dimensionless squeezing parameter $\gamma = 2$, unless otherwise stated. This corresponds to state-of-the-art squeezing parameters by $\gamma_{dB} = 10 log_10(e^{2 \gamma}) \approx 15 dB$ as reported in recent experimental implementations~\cite{Vahlbruch_2016, Liu_2025, Danilishin_2012}.\\

Continuing from Equation~\eqref{eqn:quant_payoff_q} and Equation~\eqref{eqn:expected_qi}, the EV quantum payoff $\langle\$_i^Q\rangle$ can be expressed with respect to $\textbf{x}$: 

\begin{align}
    \langle\$_i^Q\rangle &= -e^{4\gamma}\left(x_i - p_Q(\gamma) \frac{\sum_{j=1}^N x_j}{N}\right)^2 \label{eqn:quant_payoff_x}\\
    &\ p_Q(\gamma) = 1  + e^{-2\gamma}(p - 1) \label{eqn:p_Q_gamma}\ . 
\end{align}

The EV quantum payoff is now expressed in a near identical form to the classical payoff in Equation~\eqref{eqn:classical_payoff}. Most importantly, $p_Q$ in Equation~\eqref{eqn:quant_payoff_x} emerges as the effective $p$ scalar parameter in the EV quantum game, where the valid values of $p_Q$ lie between $p$ and $1$, depending on $\gamma$. Additionally, the scaling introduced by $e^{4\gamma}$ only alters the magnitude of the payoff spectrum, while the Nash equilibria and the trend of rational strategy remain unchanged. 

By taking expected values, the EV treatment brings the quantum $p$-BC closer to the original classical game, which allows the EV quantum $p$-BC to produce a definite rational behavior through an established game theoretic analysis technique. This will provide necessary insight for discussing the more complex SV quantum $p$-BC in Section \ref{sec:SV_game}. 

\subsection{Global mediation and a quantum-inspired classical game}
\label{subsec:global_mediation}

Equation~\eqref{eqn:expected_qi} reveals a global structure in the transformation of the player strategies. Defining $\lambda=e^{2\gamma}$, the expected processed bid may be written as:
\begin{align}
\langle q_i\rangle
&= \bar{x}+\lambda(x_i-\bar{x}) \nonumber\\
&= \lambda x_i+(1-\lambda)\bar{x}.
\label{eqn:global_mediation}
\end{align}

This transformation preserves the population mean, $\langle\bar{q}\rangle=\bar{x}$, while amplifying each player's deviation from that mean by the common factor $\lambda$. Equivalently, every player's bid is subject to the same global contribution $(1-\lambda)\bar{x}$, which depends on the complete strategy profile rather than on that player's strategy alone.

The same transformation can therefore be implemented without a quantum system. A classical mediator can collect the strategy profile $\mathbf{x}$, calculate the global mean $\bar{x}$, and apply Equation~\eqref{eqn:global_mediation} to every bid before evaluating the original $p$-BC payoff. This produces the same effective beauty-contest parameter: 
\begin{align}
p_Q
&=1+\lambda^{-1}(p-1) \nonumber \\
&=1+e^{-2\gamma}(p-1),
\label{eqn:global_pq}
\end{align}
and therefore leaves all subsequent game-theoretic results unchanged. Increasing the strength of the global mediation drives $p_Q$ towards unity, suppressing the iterative tendency towards boom or bust. 

\begin{figure}[ht]
    \centering
    \includegraphics[width=0.7\linewidth]{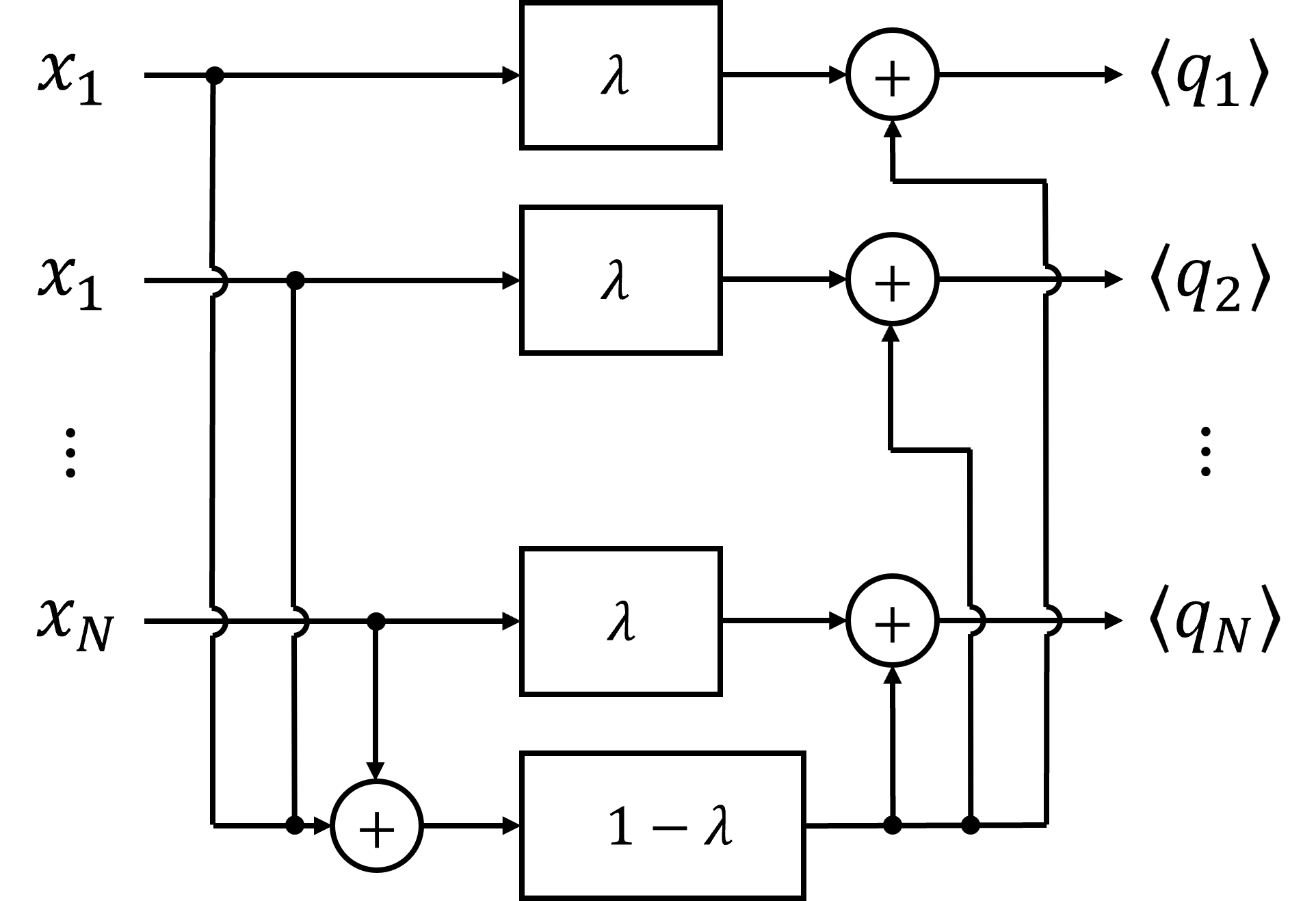}
    \caption{Logic Flow Diagram to visualize the classical computational complexity. This diagram implements the expected processed bid $\langle q_i\rangle$ in Equation~\eqref{eqn:global_mediation}, where $\lambda\in[1, \infty)$ is the common scaling factor. Each player's input bid $x_i$ passes through the multiplication operations as square boxes and the summing junctions as circled pluses to produce each player's processed bid as the output.}
    \label{fig:logic_flow}
\end{figure}

A logic flow diagram visualizing the necessary computations to obtain $\langle q_i\rangle$ is shown in Figure~(\ref{fig:logic_flow}), where the number of classical operations scales linearly as $(N-1)+(N+1)+2N \rightarrow O(N)$. Comparatively, the quantization circuit shown in Figure~(\ref{fig:quant_circuit}) also scales linearly as $O(N)$.

The quantum and quantum-inspired implementations are physically different but share the same global architecture. In the quantum circuit, multimode squeezing and the associated intermediate entanglement provide a transformation acting collectively across the player modes. In the classical construction, the analogous role is played by the global mediator through the population mean. Thus, entanglement acts as a physical implementation resource in the quantum protocol, while the underlying game-theoretic resource can be understood more generally as \emph{global mediation}: the players' strategic variables are transformed collectively rather than independently.

\subsection{Expected Iterative Elimination}\label{sec:Expected_IDSDS}
A qualitative description for the rational process of IDSDS was presented in the Section~\ref{sec:quali_iterative}. We  apply the formal mathematical process of this conventional game theoretic technique to the form of EV quantum payoff in Equation~\eqref{eqn:quant_payoff_x}.

In general, the players are presented with the strategy set $x\in[a, b]$, and the region of dominated strategy $\partial \langle\$_i^Q\rangle/\partial x_i < 0$ can be obtained by the derivative of the EV quantum payoff $\$_i$ with respect to the player input $x_i$:

\begin{align}
    \frac{\partial \langle\$_i^Q\rangle}{\partial x_i} = -2e^{4\gamma}\bigl(1-\frac{p_Q}{N}\bigr)\biggl\{(1-\frac{p_Q}{N})x_i-\frac{p_Q}{N}\sum\limits_{j\neq i}^N x_j\biggr\} \ .
\end{align}

Now starting with $x_i^{k=0} \in [a, b]$, if we iteratively derive and eliminate the $\partial \langle\$_i^Q\rangle/\partial x_i < 0$ region under every $(k-1)$-rational assumption, the expression for the $k$-rational strategy can be written as \cite{Yildiz_2012}:

\small
\begin{gather}
    x^k_i \in \left\{ \!\ 
    \begin{aligned}
        &\left[a, \left(\mathcal{I}_{\rm elim}\right)^k \cdot b\right] && \text{for } p < 1\\
        &\left[\left(\mathcal{I}_{\rm elim}\right)^k\cdot a, b\right] && \text{for } p > 1\\
        &\{(\rm x, ..., x)| x\in x_{Valid}\} &&\text{for } p = 1
    \end{aligned}\label{eqn:quantum_IDSDS}\ \right.\\
    \nonumber\\
    \mathcal{I}_{\rm elim} = \frac{p_Q(N-1)}{N-p_Q} \rightarrow \left\{ \!\ 
    \begin{aligned}
        &<1 && \text{for } p < 1\\
        &>1 && \text{for } p > 1\\
        &=1 &&\text{for } p = 1
    \end{aligned}\label{eqn:ite_param}\ \right. \ .
\end{gather}
\normalsize

The optimal $k$-rational strategies are derived for all $(k-1)$-rational assumption as shown in Equation~\eqref{eqn:quantum_IDSDS}, where the magnitude of the iterative term $\mathcal{I}_{\rm elim}$ as defined in Equation~\eqref{eqn:ite_param} ensures the lower bound does not rise above the upper bound. For a sizable player population $N\gg 1$, the iterative term converges rapidly to $p_Q$. Hence, under a large population assumption, the $k$-rational expression in Equation~\eqref{eqn:quantum_IDSDS} reduces to $x^k_i = \left[a, p_Q^k \cdot b\right]$ and aligns exactly with the described qualitative analysis of the classical Guess 2/3 of the Average example in Section~\ref{sec:quali_iterative}. Hence, the rational behavior in the EV quantum $p$-BC can be controlled by changing $p_Q$, which is achieved by adjusting the level of squeezing $\gamma$ as shown in Equation~\eqref{eqn:p_Q_gamma}.

\begin{figure}[ht]
    \centering
    \includegraphics[width=0.65\linewidth]{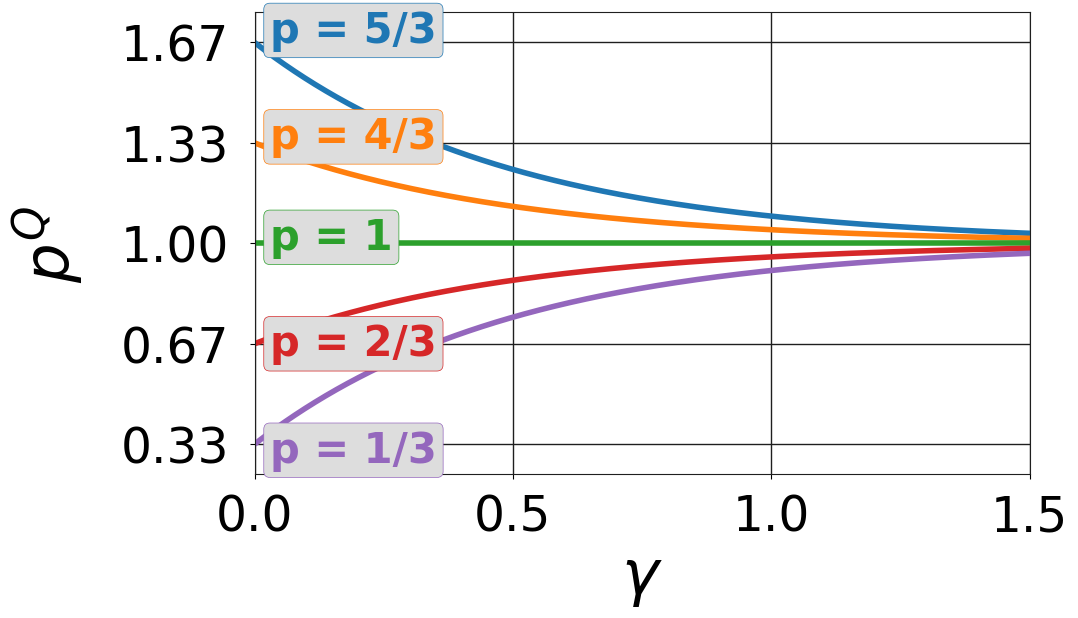}
    \caption{Effective scaling factor $p_Q$ as a function of $\gamma$: by adjusting $\gamma$ the level of squeezing, the effective $p$ in the quantum game ($p_Q$) can be adjusted towards $1$ regardless of the chosen value of $p$ in the classical game.}
    \label{fig:pQ_gamma}
\end{figure}

Under the limit of large squeezing $\gamma\gg 1$, the effective scaling parameter $p_Q$ is approximately $1$. As a result, the region of dominated strategy becomes negligible, and the strategy space eliminated through the iterative elimination process becomes insignificant. As the $k$-rational strategy becomes less relevant, the Nash equilibria approximately spans the entire strategy space $\textbf{x}_{\rm Nash} = \{(\rm x, ..., x)| x\in x_{Valid}\}$, which means boom and bust are no longer the only Nash equilibria, and the rational behavior is no longer to bid towards those extremum values. 

With the rational behavior being largely independent of $p$ under high level of squeezing, the rational behavior remains the same for a version of the EV quantum $p$-BC where each player has a separate individual scaling parameter $p_i$ in their payoff function.

\subsection{\texorpdfstring{$N$}{N}-player Quantum
\texorpdfstring{$p$}{p}-BC}

The game-theoretic analysis above has provided the trend of the rational strategies and the typical game of quantum $p$-BC. To see the game from each player's perspective and to understand how rational players may choose to interact with the rest of the players, we explore the $N$-player $p$-BC under different settings and examine the payoff spectra. 

\subsubsection{Two Players}

The $2$-player $p$-BC game can be a special case under a certain setting \cite{Grosskopf_2008}. Each of the $2$ players knows the $2/3$ average of any two numbers is always closer to the smaller number, and hence can choose $0$ to ensure an equal or higher payoff compared to the other player. However, the goal of our rational player is to maximize their own personal payoff, hence our $k$-rational strategy in Equation~\eqref{eqn:ite_param} remains unchanged for the $2$-player case. 

The $2$-player case is the simplest setting of the EV quantum $p$-BC. In Figure~(\ref{fig:2p_spec}a\&b), the player $1$'s payoff is presented as a spectrum with respect to both players' choices, and the player $2$'s payoff is identical by an exchange of the axis labels under the symmetric game $p$-BC. By examining the payoff spectra changes with increasing $\gamma$ in Figure~(\ref{fig:2p_spec}a), the most notable feature is the shifting gradient of the peak payoff curves, which are the best response curves where player $1$ bids exactly on $2/3$ of the average. These curves can be derived as: 

\begin{figure}
    \centering
    \subfloat[]{%
        \includegraphics[width=0.9\linewidth,valign=t]{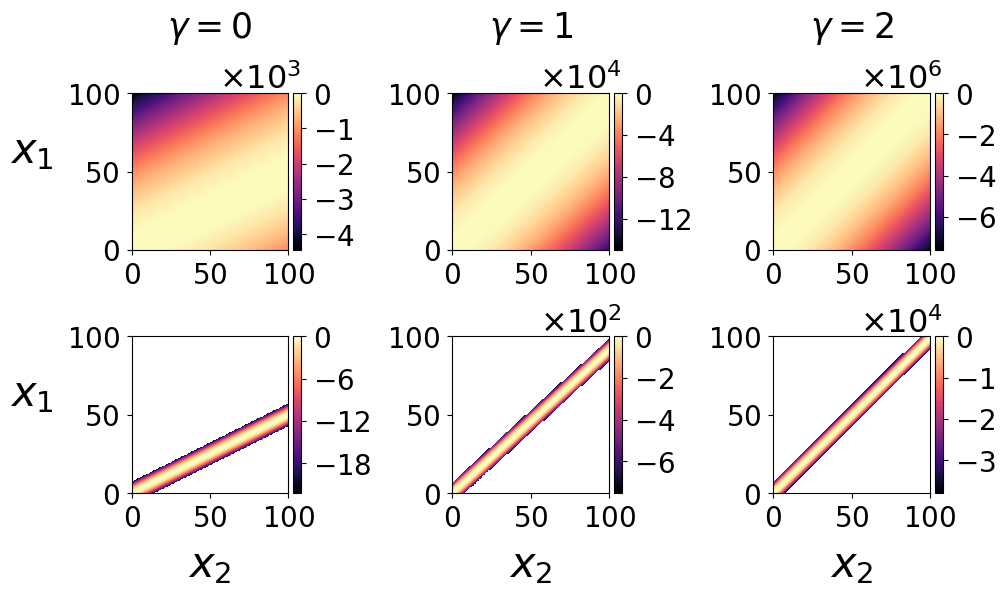}
    }\\
    \subfloat[]{%
        \includegraphics[width=0.7\linewidth,valign=t]{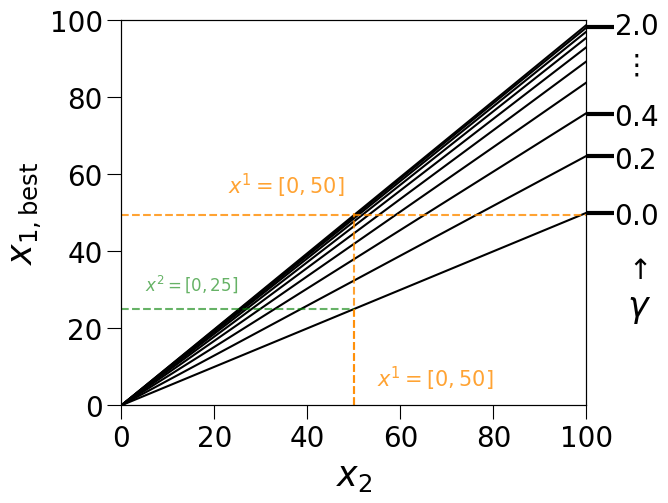}
    }
    \caption{Two-player quantum $p$-BC, player 1's payoff spectrum with respect to both players' bids: a) the payoff spectra for $\gamma$ set at $\{0,1,2\}$, where the peak gradient approaches $1$ with increasing level of squeezing; b) since the only significant feature here is the peaks of the payoff spectra, which are incidentally the best-response curves for player 1, the player interactions in the quantum game can then be simplified to observing how the best response changes with respect to $\gamma$. Additionally, the iterative elimination process is visualized by the dashed lines drawn.}
    \label{fig:2p_spec}
\end{figure}

\begin{align}
    x_{1, \rm best} &= \frac{p_Q}{2-p_Q}x_2\label{eqn:x1_2p_best}\\
    x_{2, \rm best} &= \frac{p_Q}{2-p_Q}x_1 \ .
\end{align}

The vertical extent of the best response curve in Figure~(\ref{fig:2p_spec}b) demonstrates the influence of player $2$ on player $1$'s payoff, with the bounds indicated by the intercepts for the minimum and maximum bid of player $2$. Under no entanglement $\gamma = 0$, player $1$'s optimal bid range is limited between $[0, 50]$\footnote{This result aligns with the iterative elimination process in Equation~\eqref{eqn:quantum_IDSDS}.}. With increasing level of entanglement increases $\gamma\uparrow$, player $1$'s optimal bid range grows, spanning the entire strategy space under the limit $\gamma \gg 1$. 
With larger squeezing, a player must be more sensitive to the bids of the other. 

The best response curves can also provide a visual demonstration of the iterative elimination process. For $\gamma=0$, the $1$-rational strategy is again the range of best response $x_i^{k=1} = [0,50]$, with the upper limit marked by the $x_2=100$. Assuming player $2$ follows the $1$-rational strategy, the $2$-rational strategy reduces to $x_i^{k=2} = [0, 25]$, and the process continues till the Nash equilibria bust at $k\rightarrow \infty$. Additionally, this visualization also shows the invariance of this payoff spectrum and the best response curves under the changing upper bound in the strategy space. 

\subsubsection{Three Players}

By introducing a third player, the expressions of best response curves are now:

\begin{align}
    x_{1,\rm best} &= \frac{p_Q}{3-p_Q}(x_2+x_3)\\
    x_{2,\rm best} &= \frac{p_Q}{3-p_Q}(x_1+x_3)\\
    x_{3,\rm best} &= \frac{p_Q}{3-p_Q}(x_1+x_2)\ .
\end{align}

\begin{figure}[ht]
    \centering
    \includegraphics[width=0.85\linewidth]{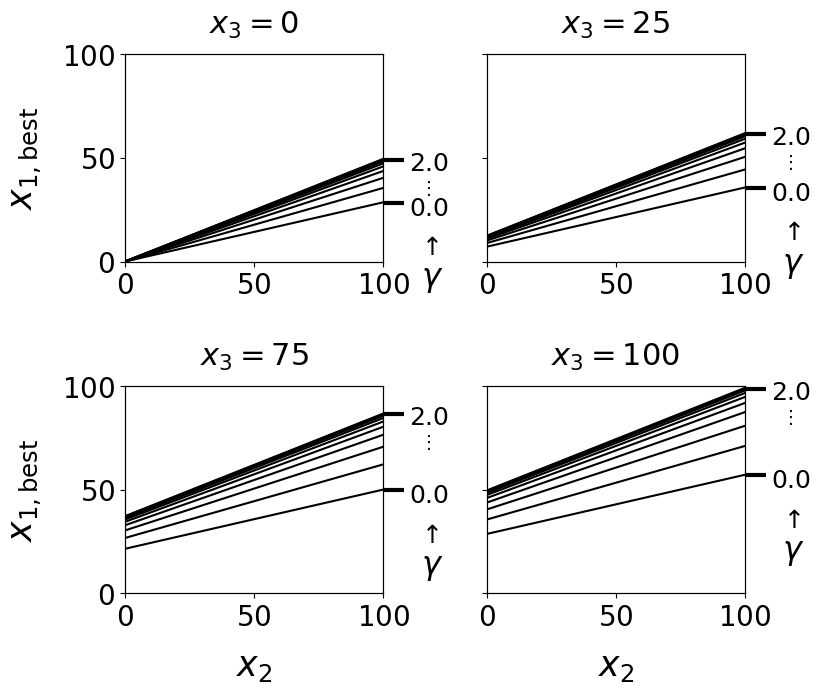}
    \caption{Three player quantum $p$-BC: Player 1's best response curves for different $\gamma$ settings given the different bids from player 2 \& 3.}
    \label{fig:3p_bare}
\end{figure}

Comparing the intercepts in Figure~(\ref{fig:3p_bare}) to Figure~(\ref{fig:2p_spec}b), player $2$ now has less influence on player $1$'s payoff as anticipated, with the span of the best response at $\gamma=2$ decreasing from $100$ to $50$. On the other hand, similar to the $2$-player game, there is an increase in responsiveness between the players under higher levels of entanglement. 

\subsubsection{Mean-field Type N-player Games}

To visualize the EV quantum $p$-BC for $N > 3$, a mean-field type approach is used by creating a group of players who always coordinate to bid the same number. Similar to the $2$-player game, the best response curves for a game of $1$ vs $N-1$ players are found with the addition of weighted average:

\begin{align}
    x_{1,\rm best} &= \frac{p_Q(N-1)}{N-p_Q}x_{N-1}\\
    x_{N-1,\rm best} &= \frac{p_Q}{N-p_Q(N-1)}x_1\label{eqn:N_1_best}\ .
\end{align}

\begin{figure}[ht]
    \centering
    \subfloat[]{\includegraphics[width=0.9\linewidth]{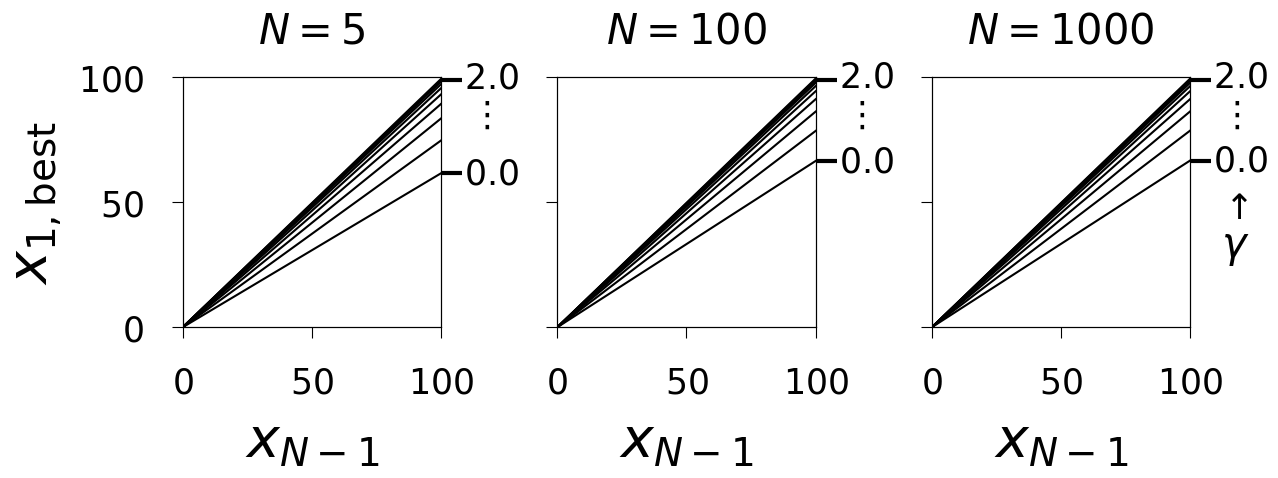}
    }\\
    \centering
    \subfloat[]{\includegraphics[width=0.6\linewidth,valign=t]{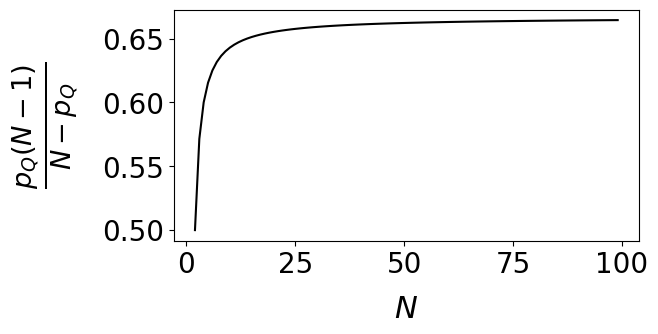}
    }\\
    \centering
    \subfloat[]{
    \includegraphics[width=0.9\linewidth]{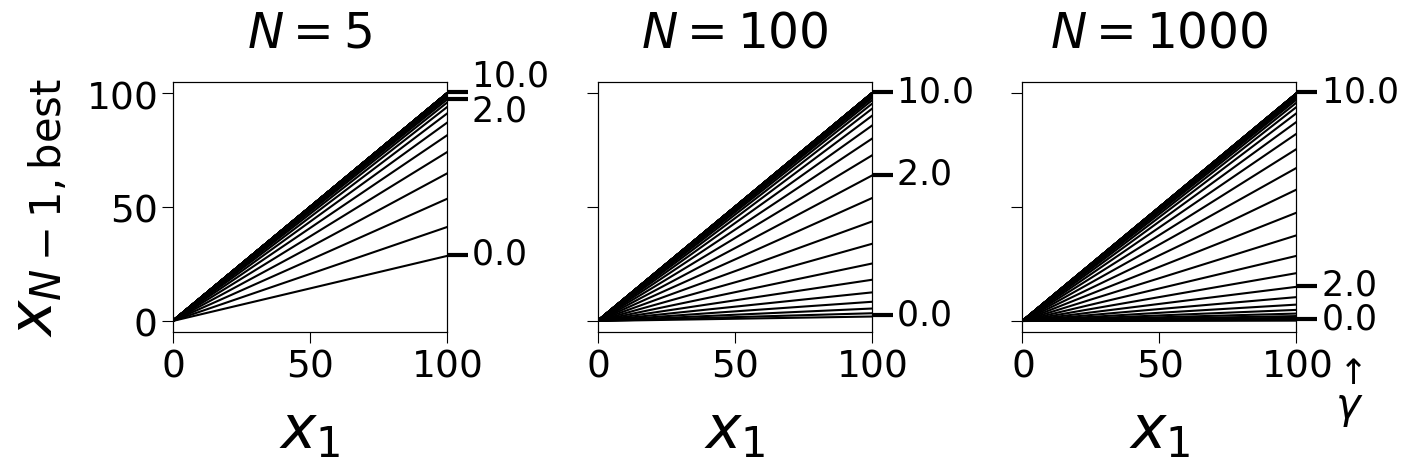}
    }\\
    \caption{Mean-field type game for $1$ vs $N-1$: a) Player 1's best response curve under different $\gamma$ settings with respect to the $N-1$ group's bid; b) The term $\frac{p_Q(N-1)}{N-p_Q}$ shows how the player should account for their bid's influence on average in calculating their bid for $2/3$ average. When $\frac{p_Q}{N-p_Q}$ is far from $p$, the player's bid has a large sway on the average bid. When $\frac{p_Q}{N-p_Q}\rightarrow p$, the average is independent of the player's bid; c) The $N-1$ group's best response curve under different $\gamma$ settings with respect to the player $1$'s bid. Here an unrealistic upper limit of $\gamma = 10$ is used to demonstrate what is required for the $N-1$ players to be influenced by the single player.}
    \label{fig:2p_mf_spec}
\end{figure}

The influence of $\gamma$ on player $1$'s best response curves is very similar to the behavior found in the $2$-player game as shown in Figure~(\ref{fig:2p_mf_spec}a) and Figure~(\ref{fig:2p_spec}b). Despite $N-1$'s significant influence on the average, player $1$ can continue to receive the maximum payoff by bidding $2/3$ of the $N-1$ group's number. The term \scriptsize$\frac{p_Q(N-1)}{N-p_Q}$\normalsize { } which accounts for player $1$'s own bid in the average quickly converges to $p_Q$ with increasing player population $N$, which agrees with player $1$'s best response curves being relatively unchanged for $N>5$.  

From the perspective of the $N-1$ group, player $1$'s bid should be insignificant due to its weighting on the average. This aligns with $N-1$ group's best response curve being horizontal under $\gamma=0$ for $N\gg 1$ in Figure~(\ref{fig:2p_mf_spec}c). However, in the limit of extreme squeezing $\gamma=10$, the best response of the $N-1$ crowd is shifted towards following the single player's bid, in the same way the single player follows the $N-1$ crowd.

Here, we present an example of the game with large squeezing $\gamma \gg 1$. Under this limit, the EV quantum $p$-BC resembles a game of Guess the Average with $p_Q = 1$, where the best response curve of the $N-1$ group in Equation~\eqref{eqn:N_1_best} becomes $x_{N-1, \rm best} = x_1$. This is independent of the player population size $N$. For a quantitative analysis, we assume a game of $N=1000$ players, where player $1$ bids $x_1=20$ and the $N-1$ group bids $x_{N-1}=10$. The $N-1$ group sees an error $0.01$, which is insignificant relative to the error of $9.99$ for player $1$. Nevertheless, the optimal strategy derived from the mathematical game-theoretic analysis suggests that the $N-1$ group will always look to improve their own payoff when possible. In this case, this error of $0.01$ from the bid $x_{N-1}=10$ decreases as $x_{N-1}\rightarrow x_1$, and hence the best response of the $N-1$ crowd is to move towards the single player's bid.

Observing the best response curves of the $N-1$ population, the magnitude of the gradients decreases as $N$ increases, reflecting the diminishing impact of the single player's bid on the average. In the limit where the single player's contribution becomes negligible, the best response for the $N-1$ population converges to $0$ for $p<1$. However, higher levels of squeezing amplify the single player's influence on the average, which lifts the best response from $0$ as shown in Figure~(\ref{fig:2p_mf_spec}c). 

Expanding to a game of $1$ vs $1$ vs $N-2$, the best response curves are found as:

\begin{align}
    x_{1,\rm best} &= \frac{p_Q(N-2)}{N-p_Q}\left(\frac{x_2}{N-2}+x_{N-2}\right)\\
    \label{eqn:3p_mf_1}
    x_{2,\rm best} &= \frac{p_Q(N-2)}{N-p_Q}\left(\frac{x_1}{N-2}+x_{N-2}\right)\\
    x_{N-2,\rm best} &= \frac{p_Q(x_1+x_2)}{N-p_Q(N-2)}\ .
    \label{eqn:3p_mf_2}
\end{align}

\begin{figure}[ht]
    \centering
    \includegraphics[width=0.85\linewidth]{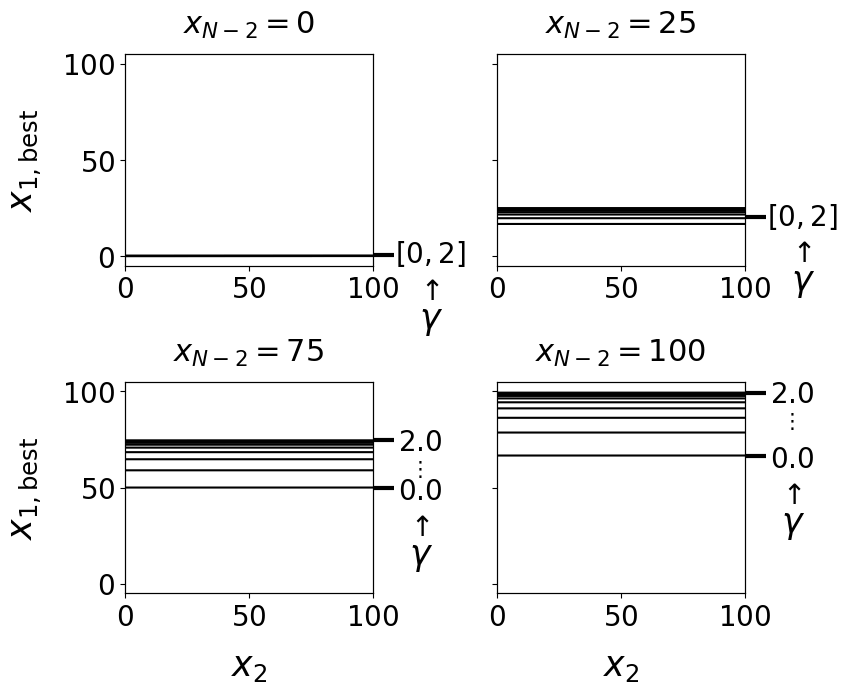}
    \caption{Mean-field type game for 1 vs 1 vs $N-2$: Player 1's best response curve given $\gamma$, player 2's bid and the $N-2$ group's bids.}
    \label{fig:3p_mf_spec}
\end{figure}

For the $2$ single players, under the large population limit, due to their increasingly insignificant contribution to the average, the $2$ players no longer have any meaningful interactions with each other as demonstrated in Figure~(\ref{fig:3p_mf_spec}). Under this setting, both players are effectively only playing the game against the $N-2$ group, where player behaviors in such interactions is reflected in the above $1$ vs $N-1$ analysis. 

Through the reparameterization of EV quantum payoff function, the conventional $k$-rational game theoretic analysis was applied and demonstrated the ability of the quantum game to disincentivize boom-and-bust behavior from rational players. Additionally, while behaviors in the classical game are shaped by players following $p$, players become more responsive to the strategy of other players in the quantum game under higher levels of entanglement.

\section{Single-Value Quantum
\texorpdfstring{$p$}{p}-Beauty Contest}
\label{sec:SV_game}

Practical applications may require the design of the quantum game to accommodate a large number of participants under heavy computational constraints. Despite the EV quantum $p$-BC's simple result, its limitation lies in the need for a large number of measurements for every participant. In this section, we present the SV quantum $p$-BC, where each player's quantum-processed bid is produced by a single quantum measurement. 

The introduced quantum probabilistic measurements can substantially alter the game's physical implementation and rational behavior. Each player's quantum-processed bid is not only influenced by the bids of other players through the entanglement, but is now also sampled from a random distribution due to the Born rule. Hence, a player may find their strategies more difficult to implement accurately, while the strategies of other players also become less predictable. This is amplified by the addition of probabilistic payoff, where each player's payoff is now dependent on the independent quantum measurements of every players' quantum-processed bids. To begin to understand the SV game and how it can shape the behaviors of rational players, we derive the form of probability distribution for the payoff function. 

In the following sections, we explore the properties and the resultant rational strategies of the SV setting of the quantized game, drawing comparisons to the EV setting described above.

\subsection{Payoff Probability Distribution}

Using the positional basis, we assume the wave function of the initial vacuum state in the quantization circuit takes the form \cite{Lvovsky_2015}: 

\begin{align}
    \nonumber \psi_0(\textbf{q}) &= \bigotimes_i^N \langle a_i|\psi_0\rangle =  \bigotimes_i^N \langle a_i|0\rangle_i 
    \label{eqn:vac_pos_wf}\\
    &= \frac{1}{\pi^{1/4}}\bigotimes_i^N \Exp{-a_i^2/2}\ .
\end{align}

The final wave function $\psi_f$ of the quantization circuit consists of separable coherent states with correlated positional displacement, where probability density function (pdf) of each measured mode $f_{q_i}$ can be expressed as:

\begin{align}
    \nonumber \psi_f (\textbf{q}, \gamma) &= \bigotimes_i^N \langle a_i|\psi_f\rangle\\
    &= \frac{1}{\pi^{1/4}}\bigotimes_i^N \expb{-(a_i-\langle q_i \rangle)^2}\\
    \nonumber f_{q_i}(a_i) &= |\langle a_i|\psi_f\rangle|^2\label{eqn:f_qi}\\
    &= \frac{1}{\pi^{1/2}}\expb{-(a_i-\langle q_i \rangle)^2}
\end{align}

\noindent where $|a_i\rangle$ refers to the positional basis, $a_i\in(-\infty, \infty)$ is the support for the random variable $q_i$, and $\langle q_i \rangle$ is as defined previously in Equation~\eqref{eqn:expected_qi}.\\

The measurement outcomes $q_i$ follow a continuous Gaussian distribution described by $f_{q_i}(a_i)$. With the quantum payoff $\$_i^Q$ described in Equation~\eqref{eqn:quant_payoff_q}, the variance of the measurements is then $\sigma^2_{\rm q_i} = 1/2$ under no squeezing $\gamma = 0$, and the expected value of each mode aligns with the input of the respective player $\langle q_i\rangle(\gamma=0) = x_i$.

\subsubsection{Payoff Distribution by Monte Carlo}
In this section, we show two separate numerical approaches to demonstrate a characteristic of the SV quantum payoff. Following the $2$-player quantum $p$-BC example from Section~\ref{subsec:Squeezing_Effect}, the two approaches will produce two different payoff spectra, which converge in the continuous limit. Both approaches begin with the Gaussian random distributions of $q_1$ and $q_2$ in Equation~\eqref{eqn:f_qi}. 

\begin{figure}[ht]
    \centering
    \subfloat[]{%
    \includegraphics[width=0.8\linewidth]{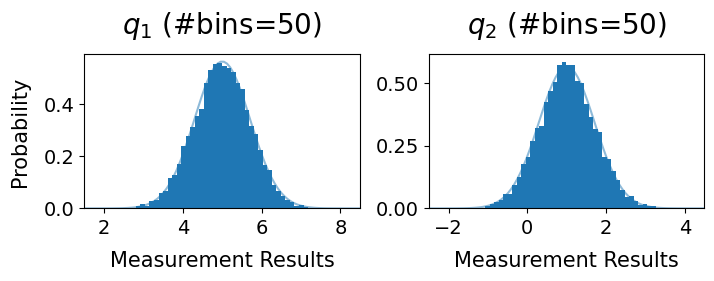}
    }\\
    \subfloat[]{%
    \includegraphics[width=0.9\linewidth]{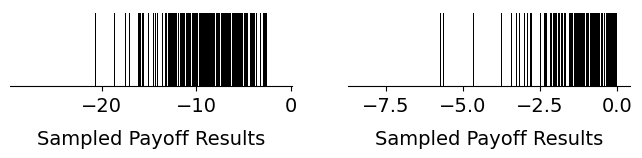}
    }\\
    
    \subfloat[]{%
    \includegraphics[width=0.8\linewidth]{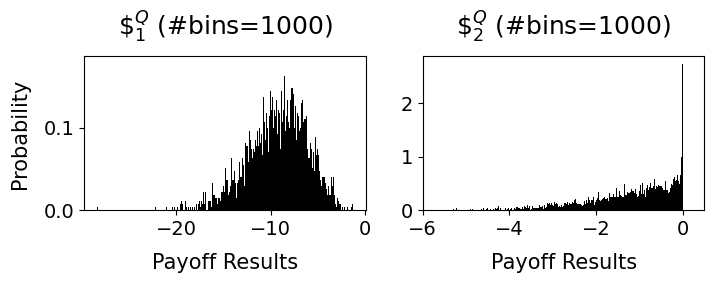}
    }\\
    
    \subfloat[]{%
        \includegraphics[width=0.8\linewidth]{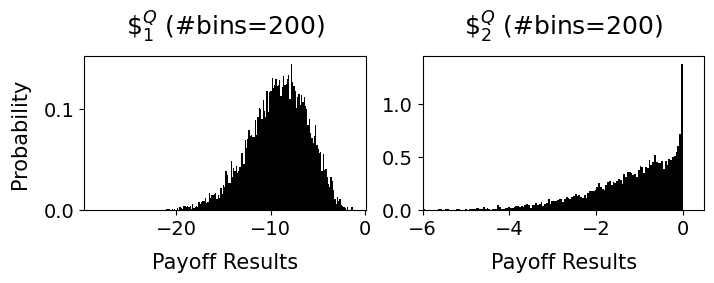}
    }\\
    \subfloat[]{%
    \includegraphics[width=0.8\linewidth]{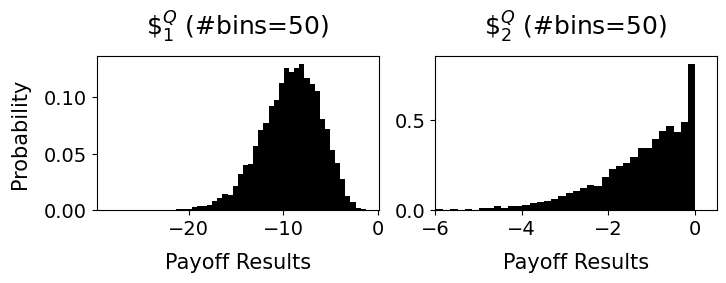}
    }
    \caption{Payoff Probability Spectrum by Monte Carlo: a) Histograms of 10,000 random samples from $q_1$ and $q_2$ each, discretized with 50 uniform bins, where the player inputs are $x_1=5$ and $x_2 = 1$; b) Calculated payoff values for both players from 10,000 pairs of $q_1$ and $q_2$; c-e) Histogram of 10,000 calculated payoff samples for both players $\$_1^Q$ \& $\$_2^Q$ with 1,000 bins, 200 bins, and 50 bins.}
    \label{fig:num_dis1}
\end{figure}

The first approach is Monte Carlo. This process is visualized in Figure~(\ref{fig:num_dis1}), where the $q_i$ distributions are randomly sampled. For each pair of measurement samples, the quantum payoff $\$_i^Q$ is computed for each player. With 10,000 samples, the probability distributions of each player's payoff can be constructed, which is presented as raw samples in Figure~(\ref{fig:num_dis1}b), and a histogram in Figure~(\ref{fig:num_dis1}c).

By observing player $2$'s payoff spectrum in Figure~(\ref{fig:num_dis1}c) with increasing number of bins, the closest bin to $0$ rapidly increases in height relative to the others, and a diverging pattern emerges at $0$ in the continuous limit. In the following, we investigate how this divergence can be introduced by processing samples of two finite probability density functions of $q_1$ and $q_2$ through the payoff function in Equation~\eqref{eqn:quant_payoff_q}.

\subsubsection{Divergence in Payoff PDF}

\begin{figure}
    \centering
    \subfloat[]{\includegraphics[width=0.85\linewidth]{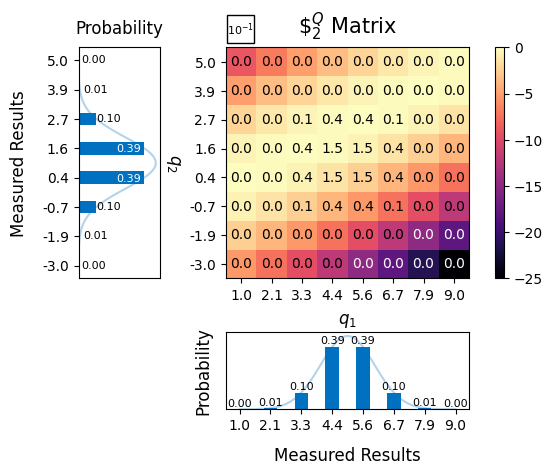}
    }\\
    
    \subfloat[]{%
    \includegraphics[width=0.7\linewidth]{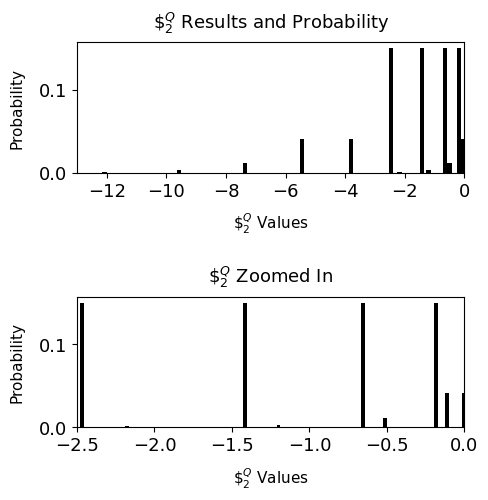}
    } \\
    
    \subfloat[]{
     \includegraphics[width=0.65\linewidth]{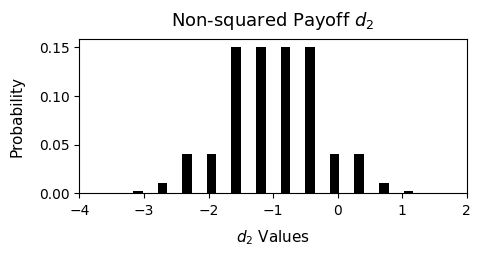}
     }
    \caption{Computing payoff distribution through uniform sampling: a) Here, the probability distributions of $q_1$ and $q_2$ are sampled at 8 distinct outcomes for each mode. The $8$ pairs of $q_1$ and $q_2$ samples give $64$ possible payoff results, each with probability given by $f_{q_1}(a_1) \times f_{q_2}(a_2)$. The payoff values are presented by a color-gradient plot, with their corresponding probability value (scaled by $\times 10^{-1}$) at the center of each colored box. b) By gathering the possible payoff results and their associated probabilities, a histogram of the payoff probability distribution is constructed. c) Similarly, the probability distribution of the bid error $d_i$ described in Equation~\eqref{eqn:non_squared_payoff} can also be built.}
    \label{fig:num_dis2}
\end{figure}

To confirm the divergence is not an artifact, we compute the spectrum through a different approach. Instead of randomly sampling $q_i$, we begin with discretizing the pdf of $q_1$ and $q_2$, and constructing the payoff matrix of player $2$ as visualized in Figure~(\ref{fig:num_dis2}a). By computing the probability for obtaining each individual payoff outcome with the pdf of $q_1$, the payoff probability spectrum is created as presented in Figure~(\ref{fig:num_dis2}b). As shown in the results of the two different numerical approaches, both have resulted in a concentrated probability near $0$ which increases rapidly with increasing number of samples and bins. This suggests that the divergence at $0$ may be an inherent feature of the pdf of the SV quantum payoff.

To investigate the origin of the divergence, we examine the component inside the quantum payoff in Equation~\eqref{eqn:quant_payoff_q}:
\begin{align}
    d_i = q_i-\frac{p}{N}\sum_j q_j \label{eqn:non_squared_payoff}
\end{align}

where $\$^Q = -d_i^2$.

The bid error $d_i$ is introduced to demonstrate the effect of squaring a random variable. The bid error represents the distance between a player's bid and $p$ of the average. The resultant pdf of $d_i$ through the numerical computation is shown in Figure~(\ref{fig:num_dis2}c), where high probability amplitudes are gathered between $-2$ and $0$. When the random variable $d_i$ is squared in the quantum payoff function, the pdf support bunches and spreads for $|d_i|<1$ and $|d_i|>1$ respectively. This behavior stems from the fact that squaring a small number $\in(0, 1)$ creates a smaller number, and squaring a large number $\in(1, \infty)$ creates a larger number. The values of $d_i$ close to zero bunch closer and closer. Amplifying the effect, the negative values of $d_i$ also switch to positive in the process of squaring. 

In the $2$-player example, the consequence of squaring a random variable is most significant in player $2$'s payoff due to their expected payoff being much closer to $0$ compared to player $1$. With the number of uniform samples for $q_1$ and $q_2$ increasing indefinitely in the continuous limit, the number of probability amplitudes around $0$ in Figure~(\ref{fig:num_dis2}b) grows at a much faster rate relative to the rest of the spectrum, which results in the divergence in the SV quantum payoff pdf. 

\subsubsection{Analytical Expression of Payoff}

To work towards a general expression for the SV quantum payoff's pdf, we return to the pdf of $q_i$ described in Equation~\eqref{eqn:f_qi}. Each player's quantum payoff involves subtracting their bid $q_i$ by the scaled summation of all player's random bid and taking the square as in Equation~\eqref{eqn:quant_payoff_q}. The subtraction and the summation of random variables can be derived through the convolution of their pdf, and the pdf of the bid error $d_i$ is found to be:

\begin{align}
    f_{d_i}(a_{d_i}) &= A\expbb{-B\left(a_{d_i} - C_i\right)^2}
    \label{eqn:qi_PDF}
\end{align}

\noindent where $a_{d_i}\in (-\infty, \infty)$ is the support for the random variable $d_i$, while $A$, $B$ and $C_{i}$ are defined as:
\begin{align}
    A &= \sqrt{\frac{N}{\pi(N-2p+p^2)}} &&\rightarrow \frac{1}{\sqrt{2\pi{\rm Var}[d_i]}}\\
    B &= \frac{N}{N-2p+p^2} &&\rightarrow \frac{1}{2{\rm Var}[d_i]}\\
    C_i &=  \langle q_i\rangle(\gamma) - \frac{p}{N} \sum^N_j \langle q_j\rangle(\gamma) &&\rightarrow {\rm E}[d_i]\ .
\end{align}

Squaring the $d_i$ involves using multiplicative convolution\footnote{For the detailed derivation, please refer to Appendix~\ref{sec:paypdf_derivation} Payoff PDF Derivation.}, and hence the pdf of the quantum payoff $\$^Q = -d_i^2$ is: 

\begin{align}
    \nonumber f_{\$_i^Q}\left(a_{\$_i^Q}\right) &= \frac{A}{2\sqrt{-a_{\$_i^Q}}} \biggl\{\expbb{-B\left(\sqrt{-a_{\$_i^Q}} - C_i\right)^2}\\
    & +\expbb{-B\left(\sqrt{-a_{\$_i^Q}} + C_i\right)^2}\biggr\} \label{eqn:payoff_pdf}\ .
\end{align}

For a simpler picture of the payoff pdf under different input bids and $\gamma$, we derive the pdf of expected value and variance:
\begin{align}
    \nonumber {\rm E}\left[\$_i^Q\right] &= \int_{-\infty}^0 a_{\$_i^Q}\cdot f_{\$_i^Q}\ da_{\$_i^Q}\\
    \nonumber &= -\left(\frac{1}{2B} + C_i^2\right)\\[1em]
    &\rightarrow -{\rm Var}\left[d_i\right] - [{\rm E}[d_i]]^2\label{eqn:payoff_ex}\\
    \nonumber\\
    \nonumber {\rm Var}\left[\$_i^Q\right] &= \int_{-\infty}^0 \left(a_{\$_i^Q}\right)^2\cdot f_{\$_i^Q}\ da_{\$_i^Q} - \left[{\rm E}\left[\$_i^Q\right]\right]^2\\
    \nonumber &= \frac{1}{B} \left(\frac{1}{2B} + 2 C_i^2\right)\\[1em]
    &\rightarrow 2[{\rm Var}[d_i]]^2 + 4{\rm Var}[d_i]\cdot [{\rm E}[d_i]]^2 \label{eqn:payoff_var}\ .
\end{align}

\begin{figure}[ht]
    \centering

    \subfloat[]{%
        \includegraphics[width=0.9\linewidth]{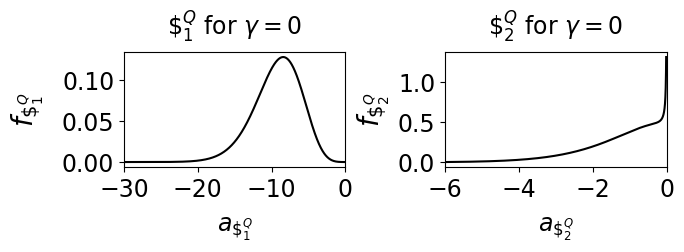}
    }\\

    \subfloat[]{%
        \includegraphics[width=0.9\linewidth]{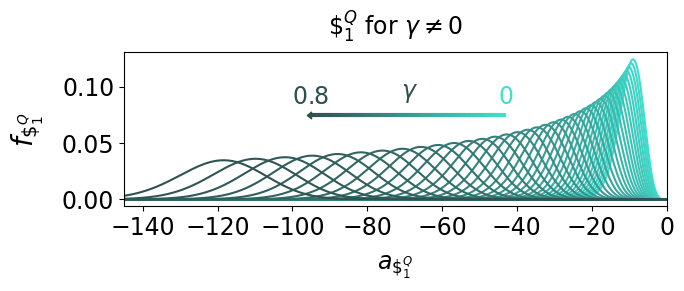}
    }\\

    \subfloat[]{%
        \includegraphics[width=0.9\linewidth]{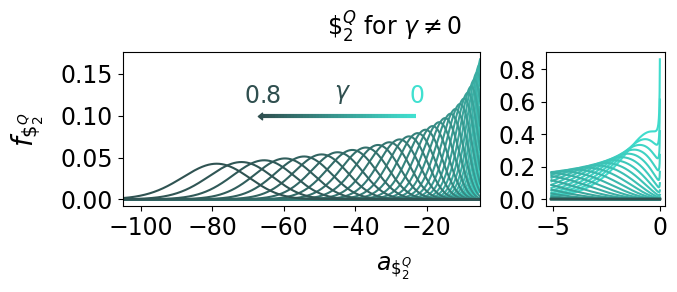}
    }\\

    \subfloat[]{%
        \includegraphics[width=0.88\linewidth]{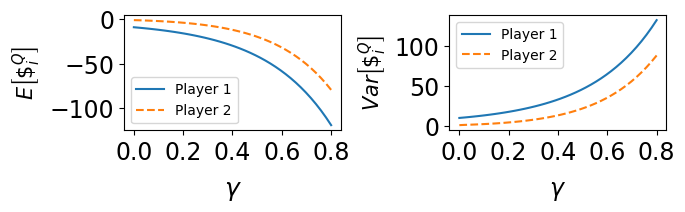}
    }

    \caption{Analytical Solution of $\$_i^Q$'s Probability Density: a) for player 1 and player 2 with support $a_{\$_1^Q}$ and $a_{\$_2^Q}$ respectively; b) $\$_1^Q$ with $\gamma\in [0, 0.8]$ (in 0.02 increments); c) $\$_2^Q$ for $\gamma\in [0, 0.8]$, where small values are plotted separately to show the divergence; d) The expected value and variance of the payoffs as a function of $\gamma$.}
    \label{fig:payoff_pdf}
\end{figure}

In the EV game, the expected payoff of each player $\langle \$_i^Q\rangle$ is determined by the expected measurement outcomes $\langle q_i\rangle$. In the SV game, we compute the payoff for each player using the processed bids $q_i$'s obtained in a  single measurement iteration, where the expected payoff ${\rm E}[\$_i^Q]$ is derived as the average of these resulting payoff values over multiple iterations. 

Under increasing level of squeezing, the variance of the payoffs increases while the expected payoff decreases for the same set of input bids. This effect can be seen by observing Equation~\eqref{eqn:payoff_ex} and \eqref{eqn:payoff_var}. The two features are also shown for the given example in Figure~(\ref{fig:payoff_pdf}b\&c). The decreasing expected payoff is a result of the squeezing operations amplifying the distance between the player's bids and the average bid as discussed in Section \ref{subsec:Squeezing_Effect}, while under lower expected payoffs, the variance increases as a result of the contribution from $\bigl(-a_{\$_i^Q}\bigr)^{-1/2}$ term in Equation~\eqref{eqn:payoff_pdf}. 

With an increasing variance and uncertainty under higher levels of squeezing, estimating the strategies of others becomes increasingly difficult. While the transient learning process of rational players can only be clearly understood through a multi-round analysis of the game, the introduction of bid uncertainty may disincentivize speculative behaviors due to the increased difficulty and significantly weakens high-order rationality strategies. 

Due to the contribution of $\bigl(-a_{\$_i^Q}\bigr)^{-1/2}$ in the payoff pdf, the divergence near $0$ becomes significant for a near $0$ expected payoff ${\rm E}\left[\$_i^Q\right]\approx0$. Given the probability is defined by the integral of some continuous interval of the pdf, the narrowness of the peak near $0$ from the divergence does not necessarily lead to the players getting better payoff on average. 

\subsection{Probabilistic Payoff \& Optimal Strategy}

The SV quantum payoff probability spectrum is the integration of its pdf function over an interval of outcomes. Given the measurement outcomes are always of finite resolution in practice, the discretization produced by the Monte Carlo simulation may be close to the observations in an implemented quantum game. Here, we demonstrate the characteristics of the SV payoff through the integration of the payoff pdf in Equation~\eqref{eqn:payoff_pdf} over equal-width intervals. 

\begin{figure}[ht]
    \centering

    \subfloat[]{%
        \includegraphics[width=0.8\linewidth]{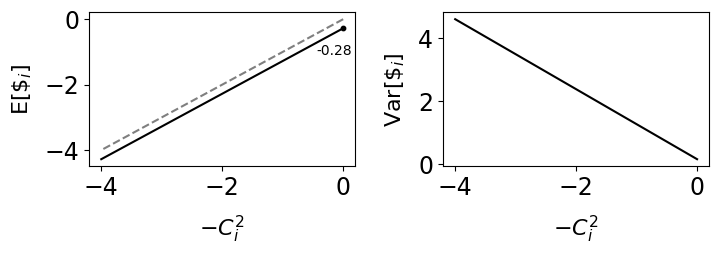}
    }\\

    \subfloat[$n_{\rm samples}=100$]{%
        \includegraphics[width=0.9\linewidth]{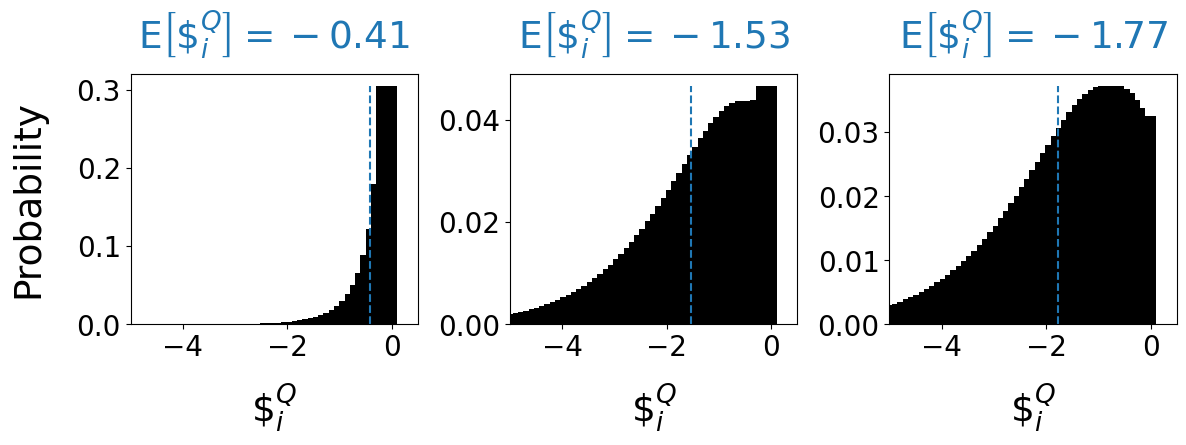}
    }\\

    \subfloat[$n_{\rm samples}=1000$]{%
        \includegraphics[width=0.9\linewidth]{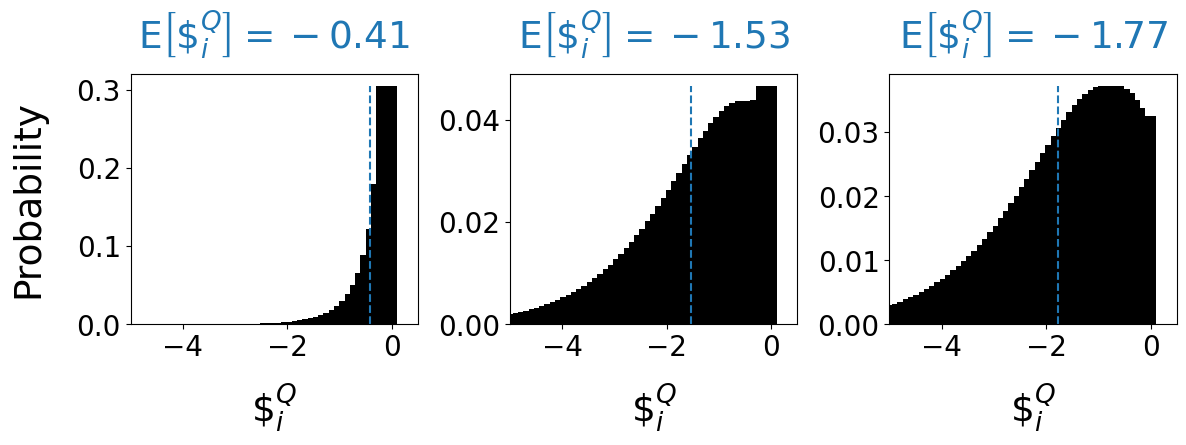}
    }

    \caption{a) Expected value and variance of the SV payoff ${\rm E}\left[\$_i^Q\right]$ \& ${\rm Var}\left[\$_i^Q\right]$ as a function of the negative squared expected bid error $-C_i^2$, with dashed lines drawn as references for gradients $-1$ and $1$. b) Resultant payoff probability spectrum given $n_{\rm samples}=100$ by adjusting the expected bid error $C_i$ and hence expected payoff ${\rm E}\left[\$_i^Q\right]$. c) Resultant payoff probability for $n_{\rm samples}=1000$.}
    \label{fig:mpo_setup}
\end{figure}

Despite the additions of probabilistic quantum measurements, the SV expected payoff ${\rm E}\left[\$_i^Q\right]$ has retained the linear dependence on $-C_i^2$ from the EV quantum payoff as proven in Equation~\eqref{eqn:payoff_ex} and visualized in Figure~(\ref{fig:mpo_setup}a). The dashed line in the figure demonstrates the equivalence between negative squared expected bid error and the EV payoff function $\langle \$_i^Q\rangle$. This is a significant result for the SV game, as it demonstrates that any rational player seeking to maximize the expected payoff in the SV quantum $p$-BC will adopt a strategy consistent with the analysis of the EV quantum $p$-BC, hence preserving the mitigation of the boom-and-bust strategy.

A rational player may also approach the game to optimize their bid for the most probable payoff outcome. For the purpose of this analysis, we define the most probable payoff for a given discretization setting $n_{\rm samples}$, using the notation $\bigl[\$^Q_i\bigr|{{\rm Max}(P_{\$^Q_i})}\bigr](n_{\rm samples})$. In any game of $p$-BC, each player's payoff is solely dependent on the inputs of the participating players. Accordingly, under fixed squeezing parameter $\gamma$, scaling parameter $p$, number of players $N$ and the number of discrete samples for pdf $n_{\rm samples}$, the expected value and the variance of payoff is only a function of the expected bid error $C_i$ and the expected measurements $\langle q_i\rangle$ as defined in Equation~\eqref{eqn:payoff_pdf}.


For a player who bids exactly on $p_Q$ of the average with an expected bid error of zero $C_i=0$, their expected payoff becomes less than $0$ ${\rm E}\left[\$_i^Q\right] = -{\rm Var}(d_i) < 0$, as demonstrated in Equation~\ref{eqn:payoff_ex}. Namely, if player $1$ guesses exactly on $2/3$ of the average in the $2$-player case, their expected payoff is ${\rm E}\left[\$_i^Q\right]\approx-0.28$ in this example instead of $0$ as in the EV quantum game. Despite the lowered expected payoff, it is important to note that the rational strategy in game theory depends on the gradient of payoff rather than the absolute value. The more significant change under the SV game is how each individual bid yields a non-zero probability of receiving the maximum payoff, as a result of the probability density diverging at $0$. 

The peak probability shifts under different discretization settings, which is largely dependent on the bid error $d_i$. By increasing the magnitude of the expected bid error $|C_i|$ from $0$, there exists a value where the most probable payoff is non-zero $\bigl[\$^Q_i\bigr|{{\rm Max}(P_{\$^Q_i})}\bigr]\neq 0$ for a given $n_{\rm samples}$, as demonstrated in Figure~(\ref{fig:mpo_setup}b) \& Figure~(\ref{fig:mpo_setup}c). We define this value as the threshold ${\rm E}\left[\$_i^Q\right] = \$_{\rm thres}^Q$.

\begin{figure}[ht]
    \centering

    \subfloat[]{%
        \includegraphics[width=\linewidth]{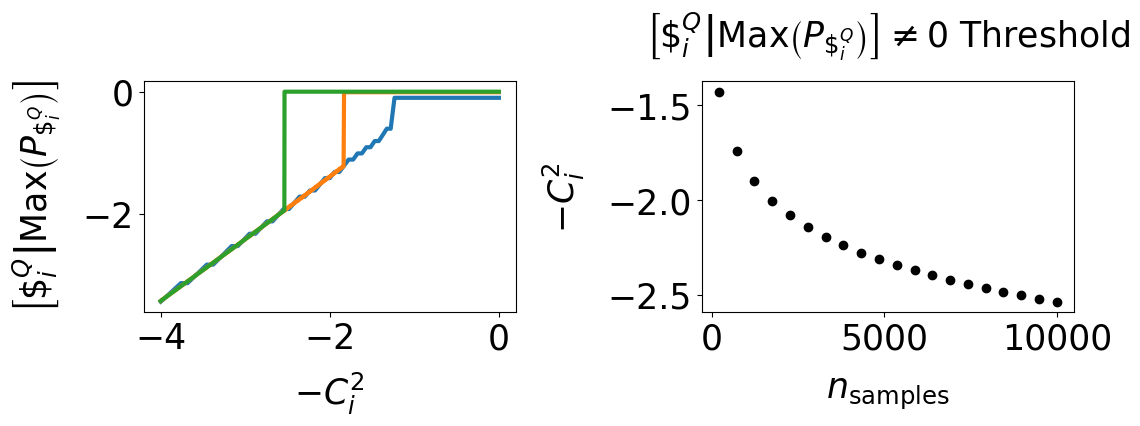}
    }\\

    \subfloat[]{%
        \includegraphics[width=0.6\linewidth]{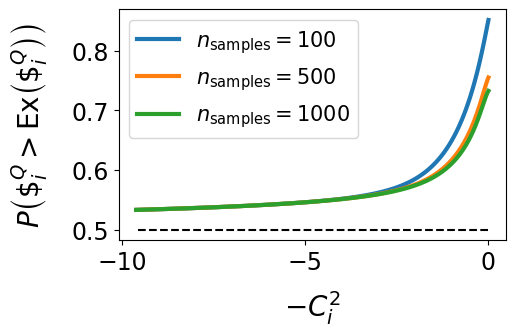}
    }
    \caption{a) Left: The most probable payoff $\bigl[\$^Q_i\bigr|{{\rm Max}(P_{\$^Q_i})}\bigr]$ as a function of negative squared expected bid error $-C_i^2$ for the discretization setting $n_{\rm samples} \in \{100, 500, 1000\}$. Right: The negative squared expected bid error threshold $-C_i^2$ for the most probable payoff $\bigl[\$^Q_i\bigr|{{\rm Max}(P_{\$^Q_i})}\bigr]\neq 0$ to be non-zero, as a function of $n_{\rm samples}$; b) Probability of obtaining better than expected payoff $\$_i^Q > {\rm E}\left[\$_i^Q\right]$ as a function of $-C_i^2$.}
    \label{fig:mpo_thres}
\end{figure}

Below the threshold, the most probable payoff \footnote{${\rm E}\left[\$_i^Q\right] < \$_{\rm thres}^Q \Rightarrow \bigl[\$^Q_i\bigr|{{\rm Max}(P_{\$^Q_i})}\bigr] = 0$.} is $0$, where the divergence term  $\bigl(-a_{\$_i^Q}\bigr)^{-1/2}$ dominates the Gaussian term of the pdf. Above the threshold, the most probable outcome approaches the expected payoff \footnote{${\rm E}\left[\$_i^Q\right] > \$_{\rm thres}^Q \Rightarrow  \bigl[\$^Q_i\bigr|{{\rm Max}(P_{\$^Q_i})}\bigr] \rightarrow {\rm E}\left[\$_i^Q\right]$.}, where the Gaussian contribution in the pdf dominates the divergence term in Equation~\eqref{eqn:payoff_pdf}. Under the limit of infinite discretization samples $n_{\rm samples} \rightarrow \infty$, the threshold diverges to negative infinity $\$_{\rm thres}^Q \rightarrow -\infty$ as demonstrated in the right figure of Figure~(\ref{fig:mpo_thres}a). Hence, there is no value where the $\bigl[\$^Q_i\bigr|{{\rm Max}(P_{\$^Q_i})}\bigr]$ becomes non-zero under the continuous limit.

Intuitively, one would expect players to benefit from the SV quantization given the non-zero probability of obtaining the maximum payoff when a player bids close to the $p$ of the average in the quantum game. To confirm this quantitatively, for each given $n_{\rm samples}$, we again shift $C_i$ and calculate the probability for a larger than expected value outcome ($\$_i^Q > {\rm E}\left[\$_i^Q\right]$) by integration. The result is as shown in Figure~(\ref{fig:mpo_thres}c). For $n_{\rm samples} \gg 1$, bidding right on the $p$ of average gives ${\rm E}\left[\$_i^Q\right] = -0.28$ with around $70\%$ probability of getting better than $-0.28$ as payoff. For worse bids under increased $|C_i|$, the chance of getting better payoff becomes the same as the chance of getting worse -- 50/50, and this is where $f_{\$_i^Q}$ becomes approximately a Gaussian, and the expected value simply becomes ${\rm E}\left[\$_i^Q\right] \approx -C_i^2$. 

Lastly, the noise introduced by the uncertainty principle is static in magnitude, and its impact on the payoff is dependent on the range and resolution of all possible bids. If the range of the displacement operation can be widened without increasing noise or decreasing bid resolution, then the pdf of the quantum measurement $q_i$ and payoff $\$_i^Q$ converges to Dirac delta functions in the limit of large strategy range ($x_i\in [a, b]$ where $b\gg a$), and hence, SV quantum $p$-BC converges to the EV game. 

Despite the added layers of complexity, the SV quantum $p$-BC approach significantly reduces the requirement on computational resources, while retaining the capability in mitigating boom-and-bust rational strategies on-average, similar to the EV quantum $p$-BC. As shown, the SV probabilistic payoff strongly rewards accurate estimations of the average bid, which may contribute to further incentivize non-speculative rational strategies.\\

\section{Conclusion}
We have demonstrated, by game theoretic analysis of the designed quantum $p$-BC, that the boom-and-bust rational strategy present in the classical $p$-BC can be contained by the quantization process. We showed the quantization scheme can relax the $k$-rationality process for the rational players, where this feature is present in both the EV and the SV setting of the quantum $p$-BC. With a chosen level of squeezing $\gamma$ and the adjustment of the quantum effective scaling parameter $p_Q$, the resistance against boom-and-bust strategies can be appropriately strengthened for a desirable level. In the last section, we presented some unique features of the quantum $p$-BC under the single-measurement setting as the SV quantum $p$-BC. We have provided quantitative insights for the SV quantum game under the measurement uncertainty, as well as the differences and similarities compared to the simpler EV game.

The SV quantum $p$-BC has the potential to create a game with similar characteristics to the EV setting while requiring significantly fewer resources. On the other hand, the complexity could lead to some unexpected exploitative strategies in the multi-round game that would alter the above described player behavior. One way to explore this can be simulation with neural network agents. This approach will also allow comparison between the learning processes of the agents in the SV and EV game, and potentially finding convergence in rational strategies as well as Nash equilibria in the SV quantum $p$-BC. 

By incorporating a designed quantization process, there is potential for a quantum market system to be built as an environment with innate resistance to boom-and-bust behavior, allowing genuine pricing of commodities to emerge naturally in a market without external intervention. Given the broad scope of the simple $p$-BC model, the quantization process has the potential to accommodate the rich complexity of a functioning economic system, and contribute to designing a sustainable market and economy.

\section{Author Contribution}

\begin{itemize}
    \raggedright
    \item \textbf{Hiu Ming Lau:} Conceptualization, Formal Analysis, Investigation, Methodology, Visualization, Writing -- Original Draft Preparation
    \item \textbf{Kareem Raslan:} Conceptualization, Investigation, Validation
    \item \textbf{Kieran Hymas:} Supervision, Writing -- review \& editing
    \item \textbf{Qiang Sun:} Investigation, Software, Supervision, Writing -- review \& editing
    \item \textbf{Azhar Iqbal:} Conceptualization, Formal Analysis
    \item \textbf{Derek Abbott:} Funding Acquisition, Methodology, Project Administration, Resources, Supervision, Validation, Writing -- review \& editing.
    \item \textbf{James Q. Quach:} Conceptualization, Methodology, Project Administration, Supervision, Validation, Writing -- review \& editing
    \item \textbf{Andrew D. Greentree:} Conceptualization, Funding Acquisition, Methodology, Project Administration, Resources, Supervision, Validation, Writing -- review \& editing
\end{itemize}

\section{Acknowledgments}

We acknowledge the joint financial support of this project from RMIT and CSIRO through the PhD program. Support from the Australian Research Council (DP240103404 and FL240100217) is gratefully acknowledged. We would like to thank Tinula Kariyasasam for supportive discussions on the Mean-Field type analysis and stylistic comments.

\bibliographystyle{apsrev4-2}
\bibliography{write_up}

\begin{widetext}
\section*{Appendix}

\appendix

\section{Quantization Circuit Wave function derivation} \label{sec:quant_wf_derivation}

The final state of the quantization circuit before measurement $|\psi_f\rangle$ as shown in Figure~(\ref{fig:quant_circuit}) can be written as:
\begin{align}
    |\psi_f\rangle &= \hat{J}^\dagger\left[\prod\limits_j^N\hat{D}_j(x_j)\right]\hat{J}|000...000\rangle \label{eqn_sup:psif} \ .
\end{align}

The expectation value of measurement result for each mode with respect to the positional basis is then:
\begin{align}
    \begin{split}
        q_i &= \langle\psi_f|\hat{X}_i|\psi_f\rangle \\
        &= \langle 000...000|\hat{J}^\dagger\left[\prod\limits_j^N\hat{D}_j^\dagger(x_j)\right]\hat{J}\hat{X}_i\hat{J}^\dagger\left[\prod\limits_j^N\hat{D}_j(x_j)\right]\hat{J}|000...000\rangle\label{eqn_sup:qi} \ .
    \end{split}
\end{align}

The form of the $N$-mode squeezing operators \cite{Lo_2003} $\hat{J}(\gamma)$ \& $\hat{J}^\dagger (\gamma)$ is chosen to be:
\begin{gather}
    \begin{aligned}
        \hat{J}(\gamma) = e^{-\gamma\hat{j}}\\
        \hat{J}^\dagger (\gamma) = e^{\gamma\hat{j}}
    \end{aligned}\\
    \nonumber \hat{j} = \sum\limits_{i,j=1}^N \left(\delta_{ij} - 1/N\right) \left(\hat{a}^\dagger_i\hat{a}^\dagger_j - \hat{a}_i\hat{a}_j\right)
    \ .
\end{gather}

With the displacement operator encoding each player's strategy:
\begin{align}
		\hat{D}_i(x_i) = \expb{-ix_i\hat{P_i}}
\end{align}

where $\hat{P}_i = i/\sqrt{2}\left(\hat{a}_i^\dagger - \hat{a}_i\right)$.\\

In the following, we derive an expression of Equation~\eqref{eqn_sup:psif} in terms of all players' inputs $x_i$ by applying Baker-Campbell-Hausdorff (BCH) to $\hat{J}^\dagger\hat{D}_k\hat{J}$ \cite{Lo_1993, Frackiewicz_2016}. We begin by applying BCH to the term $\hat{J}^\dagger\hat{a}_k\hat{J}$:

\begin{align}
    \begin{split}
        e^{\hat{A}} \hat{a}_k e^{-\hat{A}} &= \sum\limits_{m=0}^\infty \frac{1}{m!} (ad_{\hat{A}})^m(\hat{a}_k)\\
        &= \frac{1}{0!}\hat{a}_k + \frac{1}{1!}[\hat{A}, \hat{a}_k] + \frac{1}{2!}[\hat{A}, [\hat{A}, \hat{a}_k]] 
        + \frac{1}{3!}[\hat{A}, [\hat{A}, [\hat{A}, \hat{a}_k]]] + ...
    \end{split}\label{eqn_sup:BCH_sandwich}
\end{align}
where $\hat{A} = -\gamma\sum\limits_{i,j=1}^N (\delta_{ij} - 1/N) (\hat{a}^\dagger_i\hat{a}^\dagger_j - \hat{a}_i\hat{a}_j)$.\\

By expanding out the multi-layered commutators, we can find a pattern to help simplify the BCH expression:
\begin{align}
    \begin{split}
       [\hat{A}, \hat{a}_k] &= -2\gamma(- \hat{a}_k^\dagger + \frac{1}{N}\sum\limits_{l}^N \hat{a}_l^\dagger)\\
        [\hat{A}, [\hat{A}, \hat{a}_k]] &= 2\gamma [\hat{A}, \hat{a}_k^\dagger]
    \end{split}\\
    \nonumber&\text{for } m > 1:\\
    \begin{split}
        (ad_{\hat{A}})^m(\hat{a}_k) &= (2\gamma)^{m-1} [\hat{A}, \hat{a}_k^\dagger]\\
        (ad_{\hat{A}})^m(\hat{a}_k^\dagger) &= (2\gamma)^{m-1} [\hat{A}, \hat{a}_k]\ .
    \end{split}
\end{align}\\

Substitute the commutator pattern into the Equation~\eqref{eqn_sup:BCH_sandwich}, we obtain:
\begin{align}
    \begin{split}
        e^{\hat{A}} \hat{a}_k e^{-\hat{A}} &= \hat{a}_k + [\hat{A}, \hat{a}_k] + (2\gamma) \frac{1}{2!}[\hat{A}, \hat{a}_k^\dagger] + \frac{1}{3!}(2\gamma)^2 [\hat{A}, \hat{a}_k] \\
        &+ (2\gamma)^3 \frac{1}{4!}[\hat{A}, \hat{a}_k^\dagger] + (2\gamma)^4 \frac{1}{5!}[\hat{A}, \hat{a}_k] - ...\\
        &= \hat{a}_k - (2\gamma)(- \hat{a}_k^\dagger + \frac{1}{N}\sum\limits_{l}^N \hat{a}_l^\dagger) - \frac{(2\gamma)^2}{2!}(- \hat{a}_k + \frac{1}{N}\sum\limits_{l}^N \hat{a}_l)\\ 
        &- \frac{(2\gamma)^3}{3!}(- \hat{a}_k^\dagger + \frac{1}{N}\sum\limits_{l}^N \hat{a}_l^\dagger) - \frac{(2\gamma)^4}{4!}(- \hat{a}_k + \frac{1}{N}\sum\limits_{l}^N \hat{a}_l)\\
        &- \frac{(2\gamma)^5}{5!}(- \hat{a}_k^\dagger + \frac{1}{N}\sum\limits_{l}^N \hat{a}_l^\dagger) - ...\\
        &= \sum\limits_m^\infty \frac{(2\gamma)^{2m}}{(2m)!} \hat{a}_k + \sum\limits_m^\infty \frac{(2\gamma)^{2m+1}}{(2m+1)!} \hat{a}_k^\dagger\\
        &+ (1 - \sum\limits_m^\infty \frac{(2\gamma)^{2m}}{(2m)!}) \frac{1}{N}\sum\limits_{l}^N \hat{a}_l- \sum\limits_m^\infty \frac{(2\gamma)^{2m+1}}{(2m+1)!}\frac{1}{N}\sum\limits_{l}^N \hat{a}_l^\dagger\label{eqn_sup:BCH_meat}\ .
    \end{split}
\end{align}\\

The resultant summation terms in Equation~\eqref{eqn_sup:BCH_meat} are incidentally Taylor series of hyperbolic functions. Hence, the action of the squeezing operators on the annihilation and creation operators gives us: 
\begin{align}
    \begin{split}
        \hat{J}\hat{a}_k\hat{J}^\dagger &= \cosh(2\gamma) \hat{a}_k + \sinh(2\gamma) \hat{a}_k^\dagger\\
        &+ \bigl(1 - \cosh(2\gamma)\bigr) \frac{1}{N}\sum\limits_{l}^N \hat{a}_l - \sinh(2\gamma) \frac{1}{N}\sum\limits_{l}^N \hat{a}_l^\dagger
    \end{split}\\
    \begin{split}
        \hat{J}^\dagger\hat{a}_k\hat{J} &= \cosh(-2\gamma) \hat{a}_k + \sinh(-2\gamma) \hat{a}_k^\dagger\\
        &+ \bigl(1 - \cosh(-2\gamma)\bigr) \frac{1}{N}\sum\limits_{l}^N \hat{a}_l - \sinh(-2\gamma) \frac{1}{N}\sum\limits_{l}^N \hat{a}_l^\dagger\label{eqn_sup:forward_sandwich}\ .
    \end{split}
\end{align}\\

With Equation~\eqref{eqn_sup:forward_sandwich}, we can then express $\hat{J}^\dagger\hat{D}_k\hat{J}$ as:
\begin{align}
    \hat{J}^\dagger\hat{D}_k(x_k)\hat{J} &= \hat{J}^\dagger \exp\bigl\{\frac{x_i}{\sqrt{2}}(\hat{a}_k^\dagger-\hat{a}_k)\bigr\}\hat{J}\\
    \begin{split}
        &= \exp\biggl\{-i x_i \biggl[\cosh(-2\gamma)\hat{P}_k-\sinh(-2\gamma)\hat{P}_k\\
        &\qquad\quad\ + \bigl(1-\cosh(-2\gamma)\bigr)\frac{1}{N}\sum\limits_l^N \hat{P}_l+ \sinh(-2\gamma)\frac{1}{N}\sum\limits_l^N\hat{P}_l \biggr]\biggr\} \ .
    \end{split}
\end{align}\\

Hence, $|\psi_f\rangle$ can now be written as:
\begin{align}
    \begin{split}
       |\psi_f\rangle = \hat{J}^\dagger\prod_k^N\hat{D}_k(x_k)\hat{J}|0\rangle^{\otimes N} &= \bigotimes_k^N \exp\biggl\{-i \biggl[x_k \bigl(\cosh{2\gamma} + \sinh{2\gamma}\bigr)\\
        &\qquad\qquad\qquad + \bigl(1-\cosh{2\gamma}\bigr)\frac{1}{N}\sum\limits_l^N x_l - \sinh{2\gamma}\frac{1}{N}\sum\limits_l^N x_l \biggr]\hat{P}_k\biggr\}\\
        &= \bigotimes_k^N \exp\biggl\{-i\bigl[e^{2\gamma}(x_k-\bar{x})+\bar{x}\bigr]\hat{P}_k\biggr\}|0\rangle_k \label{eqn_sup:fwf}\ .
    \end{split}
\end{align}\\

For quadrature measurement on the positional basis, the expectation value is:
\begin{align}
    \begin{split}    
        \langle q_k\rangle (\gamma, \textbf{x})&= \tr{\hat{X}_k|\psi_f\rangle\langle\psi_f|} = \langle \psi_f|\hat{X}_k|\psi_f\rangle\\
        &= \langle e^{2\gamma}(x_k-\bar{x})+\bar{x}|_k\ \hat{X}_k|e^{2\gamma}(x_k-\bar{x})+\bar{x}\rangle_k\\
        &= e^{2\gamma}(x_k-\bar{x})+\bar{x}\label{eqn_sup:exqi_x}\ .
    \end{split}
\end{align}\\

Finally, we find the expected payoff as a function of all players' inputs:
\begin{align}
    \begin{split}
        \langle\$_i^Q\rangle (\gamma, \textbf{x}) &= \$_i(\langle\textbf{q}\rangle)\\
        &= -\bigl(\langle q_i\rangle - \frac{p}{N}\sum_{j=1}^N \langle q_j\rangle\bigr)^2\\
        &= -e^{4\gamma}\left\{x_i - \left[1  + e^{-2\gamma}(p - 1)\right] \bar{x}\right\}^2\\
        &= -e^{4\gamma}\left(x_i - p_Q\bar{x} \right)^2
    \end{split}
\end{align}

where $p_Q = 1  + e^{-2\gamma}(p - 1)$.

\clearpage
\section{Payoff PDF Derivation} \label{sec:paypdf_derivation}

First, we express the wave function of the initial vacuum state with respect to the positional basis as \cite{Lvovsky_2015}:

\begin{align}
    \psi_0(a_i) &= \langle a_i|\hat{D}_i(x_i)|0\rangle_1 = \frac{1}{\pi^{1/4}}\expb{-(a_i^2)/2}\label{eqn_sup:vac_pos} \ .
\end{align}
where $a_i$ is the support of mode $i$'s distribution under positional basis $\hat{X}_i$.\\

Recall Equation~\eqref{eqn_sup:fwf} and Equation~\eqref{eqn_sup:exqi_x}, the wave function before measurement is:
\begin{align}
    |\psi_f\rangle &= \expb{-i\langle q_1\rangle \hat{P}_1}|0\rangle_1 \otimes \expb{-i\langle q_2\rangle\hat{P}_2}|0\rangle_2 \otimes...\otimes\expb{-i\langle q_N\rangle\hat{P}_N}|0\rangle_N
\end{align}
where the $\gamma$ dependence in the expectation value $\langle q_i\rangle(\gamma, \textbf{x})$ is omitted to avoid clutter.\\

For some displacement $\eta_1= (\alpha_1+i\beta_1)/\sqrt{2}$, the action of our displacement operators on the vacuum state is:
\begin{align}
   \nonumber  \langle a_1|\hat{D}_1(\eta_1)|0\rangle_1 &= \langle a_1|\expb{-i \eta_1 \hat{P}_1}|0\rangle_1\\
    \nonumber &= \langle a_1|\expb{-i \alpha_1\beta_1/2}\expb{i \beta_1\hat{X}_1}\expb{-i\alpha_1\hat{P}_1}|0\rangle\\
    \nonumber &= \expb{-i \alpha_1\beta_1/2}\expb{i \beta_1\hat{X}_1}\expb{-i\alpha_1\hat{P}_1}\expb{-a_1^2/2}\\
    \nonumber &= \expb{-i \alpha_1\beta_1/2}\expb{-(a_1-\alpha_1)^2/2}\\
      \xrightarrow{\beta_1=0}&=  \expb{-(a_1-\alpha_1)^2/2}\ .
\end{align}\\

Hence, we find the final wave function in positional basis $\psi_f$, as well as the probability density function of each mode $f_{q_i}$ at the end of the quantization circuit:
\begin{align}
    \nonumber \psi_f(\langle\textbf{q}\rangle) & = \langle a_1a_2...a_N|\psi_f\rangle\\
    \nonumber &= \frac{1}{\pi^\frac{N}{4}}\expb{-(a_1-\langle q_1\rangle)^2/2}\expb{-(a_2-\langle q_2\rangle)^2/2}...\\
    &\quad\ \expb{-(a_N-\langle q_N\rangle)^2/2}\\
    \langle q_i|\psi_f\rangle &= \frac{1}{\pi^\frac{1}{4}}\expb{-(a_i-\langle q_i\rangle)^2/2}\\
    f_{q_i}(a_i)= \left|\langle a_i|\psi_f\rangle\right|^2 &= \frac{1}{\pi^{\frac{1}{2}}}\expb{-(a_i-\langle q_i\rangle)^2}\ .
\end{align}

Recall the quantum payoff function:
\begin{align}
    \nonumber \$_i^Q(\textbf{q}) &= -\left(q_i - p \frac{\sum_{j=1}^N q_j}{N}\right)^2\\
    &= -\left(\left(1-\frac{p}{N}\right)q_i - \frac{p}{N}\sum q_j^N q_j\right)^2\label{eqn_sup:quantum_payoff} \ .
\end{align}

Recall the random variable properties given $X$ and $Y$ with continuous probability density distributions:
\begin{align}
    \text{for $A=X+Y$}: \quad f_A(a) &= \int^\infty_{-\infty} f_X(\tau)f_Y(a-\tau)d\tau\\
    f_{c\cdot X}(a) &= \frac{1}{|c|}f_X\left(\frac{a}{c}\right)\\
    \text{for $A=XY$}:\quad f_A(a) &= \int^\infty_{-\infty} f_X(x)f_Y(a/x)\frac{1}{|x|}dx \ .
\end{align}

The summation term for all player bids except player $i$'s in Equation~\eqref{eqn_sup:quantum_payoff} is the convolution of $f_{q_j}(a_j)$:
\begin{align}
    \nonumber f_{\sum q_j}(a_{\sum q_j}) &= f_{q_1}(a_1)*f_{q_2}(a_2)*...*f_{q_N}(a_N) = \left(\int f_{q_1}(\tau)f_{q_2}(a_1-\tau)\ d\tau\right)*...f_{q_N}\\
    \nonumber &= \frac{1}{\pi^\frac{2}{2}}\left(\int \expb{-(\tau-\langle q_1\rangle)^2}\expb{-(a_2-\tau-\langle q_2\rangle)^2}\ d\tau\right)*...f_{q_N}\\
    \nonumber &= \frac{1}{\pi}\sqrt{\frac{\pi}{2}}\left(\expb{-\frac{1}{2}(a_{1*2} - \langle q_1\rangle - \langle q_2\rangle)^2 }\right)*f_{q_3}*...*f_{q_N}\\
    \nonumber &= \frac{1}{\pi}\sqrt{\frac{\pi}{2}}\frac{1}{\sqrt{\pi}}\sqrt{\frac{2\pi}{3}}\left(\expb{-\frac{1}{3}(a_{1*2*3} - \sum^3_{j} \langle q_j\rangle)^2}\right)*f_{q_4}*...*f_{q_N}\\
    &= \frac{1}{\sqrt{(N-1)\pi}}\expb{-\frac{1}{N-1}(a_{\sum q_j} - \sum^N_{j\neq i} \langle q_j\rangle)^2} \ .
\end{align}

Then for the bid error $d_i = (1-p/N)q_i-p/N\sum q_j$, its probability distribution is:

\begin{align}
    \nonumber f_{d_i}(a_{d_i}) &= \frac{1}{|1-p/N|}f_{q_i}(a_i/(1-p/N))*\frac{1}{|-p/N|}f_{\sum q_j}(a_{\sum q_j}/(-p/N))\\
    \nonumber &= \frac{1}{1-p/N}\left(\frac{1}{\sqrt{\pi}}\expb{-(a_i/(1-p/N)-\langle q_i\rangle)^2}\right)*\\
    \nonumber &\quad\ \frac{1}{p/N}\left(\frac{1}{\sqrt{(N-1)\pi}}\expb{-\frac{1}{N-1}(a_{\sum q_j}/(-p/N) - \sum^N_{j\neq i} \langle q_j\rangle)^2}\right)\\
    \nonumber &= \frac{N}{N-p}\left(\frac{1}{\sqrt{\pi}}\expb{-\frac{N^2}{(N-p)^2}(a_i-(1-p/N)\langle q_i\rangle)^2}\right)*\\
    \nonumber &\quad\ \frac{1}{p/N}\left(\frac{1}{\sqrt{(N-1)\pi}}\expb{\frac{1}{N-1}\frac{N^2}{p^2}(a_{\sum q_j} - (-p/N)\sum^N_{j\neq i} \langle q_j\rangle)^2}\right)\\
    \nonumber &\quad\ \text{Given Gaussian distributions, take the sum of the means, sum of variances}\\
    \nonumber &= \sqrt{\frac{N}{\pi(N-2p+p^2)}}\cdot\expb{-\frac{N}{N-2p+p^2}\left( a_{d_i} - \left[(1-p/N)\langle q_i\rangle - p/N \sum^N_{j\neq i} \langle q_j\rangle\right]\right)^2}\\
    &= A \expb{-B\left(a_{d_i} - C_i\right)^2}
\end{align}

where:
\begin{align*}
    A &= \sqrt{\frac{N}{\pi(N-2p+p^2)}} = \sqrt{\frac{B}{\pi}}\\
    B &= \frac{N}{N-2p+p^2}\\
    C_i &=  \langle q_i\rangle - p \sum^N_j \langle q_j\rangle\ .
\end{align*}

Finally, to find the pdf of the random quantum payoff $\$^Q$, we square the random bid error $d_i$ and find its pdf by \cite{Watkins_2009}:

\begin{align}
    Y &= X^2\\
    \nonumber F_Y(y) &= P(Y\le y) = P(X^2 \le y) = P(|X|\le y)\\
    \nonumber &= P(-\sqrt{y}\le X\le\sqrt{y}) = F_X(\sqrt{y})-F_X(-\sqrt{y})\\
    f_Y(y) &= \frac{dF_Y(y)}{dy} = \frac{1}{2\sqrt{y}}f_X(\sqrt{y})+\frac{1}{2\sqrt{y}}f_X(-\sqrt{y})\ .
\end{align}

Hence finally, the analytical expression for the payoff probability distribution in our continuous-variable quantization can be found as:
\begin{align}
    f_{-\$_i^Q}(a_{\$_i^Q}) &= \frac{A}{2\sqrt{a_{\$_i^Q}}} \left(\expb{-B\left(\sqrt{a_{\$_i^Q}} - C_i\right)^2}+\expb{-B\left(\sqrt{a_{\$_i^Q}} + C_i\right)^2}\right)\\
    f_{\$_i^Q}(a_{\$_i^Q}) &= \frac{A}{2\sqrt{-a_{\$_i^Q}}} \left(\expb{-B\left(\sqrt{-a_{\$_i^Q}} - C_i\right)^2}+\expb{-B\left(\sqrt{-a_{\$_i^Q}} + C_i\right)^2}\right)\ .
\end{align}

The expressions for expected value and variance can be obtained by symbolic computation:
\begin{align}
    \nonumber {\rm E}\left[\$_i^Q\right] &= \int_{-\infty}^0 a_{\$_i^Q}\cdot f_{\$_i^Q}\ da_{\$_i^Q}\\
    \nonumber &= \frac{A \sqrt{\pi}}{2 B^{3/2}}\left(-1 - 2 B C_i^2\right)\\
    \nonumber&\text{for $A = \sqrt{\frac{B}{\pi}}$:}\\
    &= -\sqrt{\frac{B}{\pi}}\frac{\sqrt{\pi}}{2 B^{3/2}}2B\left(\frac{1}{2B} + C_i^2\right)\\
    &= -\left(\frac{1}{2B} + C_i^2\right)\\
    &= -{\rm Var}\left[d_i\right] - [{\rm E}[d_i]]^2\\
    \nonumber {\rm Var}\left[\$_i^Q\right] &= \int_{-\infty}^0 \left(a_{\$_i^Q}\right)^2\cdot f_{\$_i^Q}\ da_{\$_i^Q} - \left({\rm E}\left[\$_i^Q\right]\right)^2\\
    \nonumber &= \frac{1}{4 B^3}\left(A \sqrt{B} \left(3 + 4 B C_i^2 (3 + B C_i^2)\right)\sqrt{\pi} - \left(A + 2 A B C_i^2\right)^2 \pi\right)\\
    \nonumber&= \frac{A}{4 B^3}\left((3\sqrt{B\pi}-A \pi) + 4B(3\sqrt{B\pi}- A\pi) C_i^2  + 4 B^2(\sqrt{B\pi}-A \pi) C_i^4\right)\\
    \nonumber&\text{for $A = \sqrt{\frac{B}{\pi}}$ \& $A\pi = \sqrt{B\pi}$:}\\
    &= \sqrt{\frac{B}{\pi}}\frac{1}{4 B^3}\left(2\sqrt{B\pi} + 8B\sqrt{B\pi} C_i^2\right)\\
    &= \frac{1}{4 B^3} 4B^2\left(\frac{1}{2B} + 2 C_i^2\right)\\
    &= 2[{\rm Var}[d_i]]^2 + 4{\rm Var}[d_i]\cdot [{\rm E}[d_i]]^2\ .
\end{align}

\end{widetext}

\end{document}